\documentclass[]{emulateapj}
\usepackage{epsfig, natbib, graphicx, color,bm}

\input epsf 

\usepackage{color}

\newcommand{\tablecomment}[1]{\let\thefootnote\relax\footnotetext{#1}}

\newcommand{\Mpc}{\mbox{Mpc}}

\newcommand{\msun}{M_\odot}

\newcommand{\bmm}[1]{\mathbf{#1}}

\newcommand{\nn}{\nonumber}

\newcommand{\fgas}{f_{\rm gas}}

\newcommand{\avg}[1]{\left\langle #1 \right\rangle}

\newcommand{\Mf}{M_{500}}

\newcommand{\bC}{\bmm{C}}

\newcommand{\ergs}{\mbox{ergs}}

\newcommand{\be}{\begin{equation}}
\newcommand{\ee}{\end{equation}}
\newcommand{\bea}{\begin{eqnarray}}
\newcommand{\eea}{\end{eqnarray}}

\newcommand{\Nt}{N_{200}}

\newcommand{\keV}{\mbox{keV}}
\newcommand{\Ysz}{Y_{\rm SZ}}
\newcommand{\Yx}{Y_{\rm X}}
\newcommand{\Tx}{T_{\rm X}}
\newcommand{\Lx}{L_{\rm X}}

\newcommand{\Cov}{\mbox{Cov}}

\newcommand{\rosat}{{\it ROSAT}}
\newcommand{\chandra}{{\it Chandra}}
\newcommand{\xmm}{{\it XMM}}
\newcommand{\planck}{{\it Planck}}

\newcommand{\Mgas}{M_{\rm gas}}

\newcommand{\Rf}{R_{500}}

\newcommand{\ysz}{y_{sz}}
\newcommand{\bS}{\bmm{S}}
\newcommand{\bs}{\bmm{s}}
\newcommand{\balpha}{\bm{\alpha}}

\newcommand \sigmuone {\sigma_{\mu 1}}
\newcommand \sigmutwo {\sigma_{\mu 2}}

\shortauthors{Rozo et al.}
\shorttitle{A Comparative Study of Local Galaxy Clusters: II: X-ray and SZ Scaling Relations}

\begin{document}
\title{A Comparative Study of Local Galaxy Clusters: II: X-ray and SZ Scaling Relations}
\author{Eduardo Rozo\altaffilmark{1,2}, August E. Evrard\altaffilmark{3}, Eli S. Rykoff\altaffilmark{4,5}. James G. Bartlett\altaffilmark{6,7}, }
\altaffiltext{1}{Einstein Fellow, Department of Astronomy \& Astrophysics, The University of Chicago, Chicago, IL 60637.}
\altaffiltext{2}{Kavli Institute for Cosmological Physics, Chicago, IL 60637.}
\altaffiltext{3}{Departments of Physics and Astronomy and Michigan Center for Theoretical Physics, University of Michigan, Ann Arbor, MI 48109.}
\altaffiltext{4}{SLAC National Accelerator Laboratory, Menlo Park, CA 94025.}
\altaffiltext{5}{Lawrence Berkeley National Laboratory, Berkeley, CA 94720.}
\altaffiltext{6}{APC, AstroParticule et Cosmologie, Universit\'e Paris Diderot, CNRS/IN2P3, CEA/lrfu, Observatoire de Paris, Sorbonne Paris
Cit\'e, 10, rue Alice Domon et L\'eonie Duquet, Paris Cedex 13, France.}
\altaffiltext{7}{Jet Propulsion Laboratory, California Institute of Technology, 4800 Oak Grove Drive, Pasadena, CA, U.S.A.}

\begin{abstract}
We compare cluster scaling relations published for three different samples selected via X-ray and Sunyaev-Zel'dovich (SZ) signatures.   
We find tensions driven mainly by two factors:  i) systematic differences in the X-ray cluster observables used to derive the  
scaling relations, and ii) uncertainty in the modeling of how the gas mass of galaxy clusters 
scales with total mass.   All scaling relations are in agreement 
after accounting for these two effects.   We describe
a multivariate scaling model that enables a fully self-consistent treatment of multiple observational catalogs in the presence of property covariance,
and apply this formalism when interpreting published results.  The corrections due to scatter and observable covariance can be
significant.  For instance, our predicted $\Ysz$--$L_X$ scaling relation differs from that derived using the naive ``plug in'' method
by $\approx 25\%$.  Finally, we test the mass normalization
for each of the X-ray data sets we consider by applying a space density consistency test: we compare the observed REFLEX luminosity 
function to expectations from published $\Lx$--$M$ relations convolved with the mass function for a WMAP7 flat $\Lambda$CDM model.  
\end{abstract}

\section{Introduction}

The study of scaling relations of galaxy clusters is one of significant importance, both from a cosmological
and an astrophysical point of view.  Cosmologically, mass--observable scaling relations are a fundamental 
component of all work that exploits the abundance of galaxy clusters for 
constraining cosmological parameters \citep[e.g.][]{henryetal09, vikhlininetal09, mantzetal10a, rozoetal10a}.
Astrophysically, the existence of a self-similar model \citep{kaiser86,bohringeretal12} allows one to use departures from this
self-similar expectation as a probe of feedback and non-thermal processes in cluster and galaxy 
formation \citep[e.g.][]{rowleyetal04,magliocchettibruggen07,mantzetal10b,mittaletal11,maughanetal12,eckmilleretal11}.  
Likewise, the scatter of galaxy clusters about the scaling relations can also be a direct probe of 
the state of the intra-cluster gas (e.g. the presence or absence of a cool core) and/or dynamical 
state \citep[e.g.][]{fabjanetal94,hartleyetal08,rasiaetal11,krauseetal12}.
Indeed, these ideas have spurred large suites
of simulations in which the physics contributing to the state of the intra-cluster gas are systematically varied,
both to give a range of plausible evolution models, and to help guide theoretical interpretations of
current observations \citep[e.g.][]{nagai06,nagaietal07b,staneketal10,shortetal10,battagliaetal11,fabjanetal11,kayetal12}.

All this work, however, is critically dependent on our ability to robustly measure empirical  
cluster scaling relations.
Here, we take a pragmatic approach to estimating the level of systematic differences in observed X-ray and SZ scaling
relation by comparing the published values of three different sets of works:
a \planck\ and \xmm\ based analysis 
\citep{prattetal09,arnaudetal10,planck11_local}, henceforth referred
simply as P11-LS,
 a \chandra\ and \planck\ 
based analysis \citep{vikhlininetal09,rozoetal12a}, henceforth referred
to as V09, and a second, independent
\chandra\ based analysis \citep{mantzetal10b}, henceforth referred to as M10.
As is demonstrated below, these three works are often in tension with one another.  
Understanding the origins of these tensions is of paramount importance for validating the use of galaxy clusters as probes of precision
cosmology.

This is the second in a series of papers that aims to perform a detailed study of local cluster scaling relations
from X-ray and SZ selected clusters catalogs.  In the first paper --- \citet{rozoetal12b}, henceforth
paper I --- we demonstrate significant systematic differences between
different groups on raw cluster observables, including X-ray luminosity, $\Lx$, temperature, $\Tx$, and gas thermal energy, $\Yx$.  
More importantly, hydrostatic mass estimates vary substantially.  

Paper I characterizes relative offsets in these quantities, and that information forms the foundation for the current work.  Indeed,
we demonstrate below that the tension in the scaling relations between the three 
works we consider is ultimately sourced by the systematic differences in cluster observables 
identified in paper I.

Having fully understood the origin of the tension between the three groups,
we then proceed to derive a self-consistent set of multi-variate
scaling relations from each of these data sets \citep[see also][]{whiteetal10}.  
By ``self-consistent'', we mean
that our propagated scaling relations are explicitly derived from a probabilistic model
that account for the scatter and possible covariance between clusters observables,
arising for instance from local large scale structure \citep[e.g.][]{whiteetal10,nohcohn11,nohcohn12,anguloetal12}
Most work to date relies on simple ``plug-in'' methods to propagate scaling relations.
That is, given two scaling relations $Y(X)$ and $Z(Y)$, the $Z$--$X$ scaling relation
is assumed to take the form $Z(Y(X))$.  
Recent work with numerical simulations and/or Monte Carlo analyses have made it clear, however, that
this naive method is generically biased, with the error depending on 
both the scatter and covariance between the observables at 
hand \citep{rozoetal09a,whiteetal10,biesiadzinskietal12,anguloetal12}.
The two specific cases we consider are using the $M$--$\Yx$
and $\Ysz$--$\Yx$ scaling relations to derive the $\Ysz$--$M$ scaling relation, and using
the $\Lx$--$M$ and $\Ysz$--$M$ scaling relations to derive the $\Ysz$--$\Lx$ scaling
relation.      The latter example is one where 
ignoring scatter corrections can result in biases as large as $20\%-30\%$.

Having derived the cluster scaling relations for each of the three data sets we consider,
and having identified the origin of the differences them, we turn to investigate which
of these analyses, if any, is consistent with cosmological expectations, for 
a WMAP7+BAO+$H_0$ flat $\Lambda$CDM
best fit cosmological model of \citet{komatsuetal11}.  Specifically, we consider each
of the published $L_X$--$M$ scaling relation in turn, convolve them with the \citet{tinkeretal08}
mass function appropriate for the afformentioned cosmology, and then compare the resulting
predicted abundance with the X-ray luminosity function from the REXCESS 
catalog \citep{bohringeretal02,bohringeretal04}.

In addition to being interested in the X-ray and SZ scaling
relations of galaxy clusters in their own right, one of our main motivations for pursuing this work  
is to take a closer look at the recent results by \citet{planck11_optical}.  In that work, it was found 
that the $\Ysz$--$N_{200}$ scaling relation of optically-identified maxBCG clusters \citep{koesteretal07a} were
inconsistent with predictions derived from a combination of optical \citep{rozoetal09a} and X-ray data sets \citep{arnaudetal10}.   
Because these predictions depend on the X-ray and SZ
scaling relations of galaxy clusters, it is imperative that systematic uncertainties associated with these predictions be quantified, 
and that the scaling relations be self-consistently propagated in the presence of non-zero scatter.  The results from this comparison
will be presented in a subsequent paper, Rozo et al. (in preparation), henceforth referred to simply
as paper III.  For the purposes of this work, this means
we will focus our attention on cluster
scaling relations at $z=0.23$ exclusively, the median redshift of maxBCG clusters.

The layout of the paper is as follows.  In \S~\ref{sec:data} we describe
the data used in this work.  This data takes the form of the $\Lx$--$M$, 
$M$--$\Yx$, and $\Ysz$--$\Yx$ scaling relations derived from each
of the data sets we consider.  
Section \ref{sec:cluster_relations} compares these three scaling relations
to each other, and demonstrates that after correcting for the
systematic offsets in X-ray observables identified in paper I --- and tilting the M10
relations to an $\fgas\propto M^{0.15}$ model --- all scaling
relations are in good agreement with each other.
We then turn to self-consistently
deriving the $\Ysz$--$M$ (section \ref{sec:yszm}) and $\Ysz$--$\Lx$ (section \ref{sec:yszlx})
scaling relations from the input scaling relations summarized in section \ref{sec:data}.
Section \ref{sec:yszlx} also explicitly compares our self-consistently derived scaling relation
with the data from \citet{planck11_local}. 
Section \ref{sec:cosmology} tests whether the scaling relations
from each of the three works are consistent with the observed cluster luminosity
function in the current best-fit $\Lambda$CDM cosmology.  Section \ref{sec:summary}
presents a summary of our results.

In all cases, scaling relation parameters are computed assuming a flat $\Lambda$CDM cosmology
with $\Omega_m=0.3$ and $h=0.7$.  When comparing abundances to cosmological expectations
for a WMAP7+BAO+$H_0$ cosmology, we use the best fit model from \citet{komatsuetal11} ---
which has $\Omega_m=0.275$, $h=0.702$, $n_s=0.968$, and $\sigma_8=0.816$ --- to compute
the predicted halo mass function.  As discussed
in section \ref{sec:cosmology}, the scaling relations we use are still those that were computed
assuming $\Omega_m=0.3$.  This is slightly inconsistent, but is necessary in the absence of precise knowledge of the
degeneracies between each set of scaling relation parameters and cosmological parameters.  We note, however,
that because the changes in cosmological
distances to low redshifts are very mild between the two cosmological models, 
we expect this inconsistency will only
impact our predictions at the few percent level at most.  This level of uncertainty is not enough to modify our conclusions.

Unless otherwise noted, all total masses, $M$, employ the $\Mf$ convention, where $\Mf$ is defined as the mass within a radius, $\Rf$, that encompasses a mean interior density of $500$ times the critical density of the universe, $\rho_c(z) = 3 H^2(z)/8\pi G$.  


\section{Data}
\label{sec:data}


\begin{deluxetable*}{lllllll}
\tablewidth{0pt}
\tablecaption{Input Cluster Scaling Relations at $z=0.23$}
\tablecomment{In all cases, we assume the $\psi$--$\chi$ relation takes the form
$\avg{\ln \psi}= a +\alpha \ln(\chi/\chi_0)$.  Our choice of units are $10^{14}\ \msun$ for mass, 
$10^{44}\ \mbox{ergs/s}$ for $\Lx$, $10^{14}\ \msun \keV$ for $\Yx$, and 
$10^{-5}\ \Mpc^2$ for $D_A^2\Ysz$ and $C\Yx$.  Unless otherwise
noted, we set $\chi_0$ to the reference scale in the cited work.  All scaling relations are evaluated at
$z=0.23$, the median redshift of the maxBCG cluster sample.}
\tablehead{
Relation & $\chi_0$ & $a$ & $\alpha$ & $\sigma_{\ln \psi|\chi}$ & Citation & Data Set}
\startdata
$\Lx$--$\Mf$ & 4.8 & $1.16 \pm 0.09$ & $1.61\pm 0.14$ & $0.396\pm 0.039$ & \citet{vikhlininetal09} & V09\\ 
$\Lx$--$\Mf$ & 2.0 & $0.08 \pm 0.08$ & $1.62\pm 0.11$ & $0.411\pm 0.070$ & \citet{prattetal09} & P11-LS \\
$\Lx$--$\Mf$ & 10.0 & $2.11\pm 0.18$ & $1.34\pm 0.05$ & $0.414\pm 0.044$& \citet{mantzetal10b} & M10 \\
$\Lx$--$\Mf$ & 4.0 & $0.98$ & $1.52$ & --- & Reference & --- \\
\hline
\hline
$\Mf$--$\Yx$ & 3.0 & $1.53\pm 0.04$ & $0.57\pm 0.03$ & $\leq 0.07$\footnote{\citet{vikhlininetal09} only
state that the scatter is undetectable given the errors on hydrostatic mass estimates, but that
this is consistent with $7\%$ scatter as predicted by \citet{kravtsovetal06}.  We implement this scatter in our
analysis as a uniform prior in the variance with the maximum value quoted above.} & \citet{vikhlininetal09} & V09 \\
$\Mf$--$\Yx$ & 2.0 & $1.23\pm 0.02$ & $0.56\pm 0.02$ & $\leq 0.09$\footnote{\citet{arnaudetal07}
quote a scatter of $0.087$, but provide no error bars.  We implement this scatter in our analysis as a uniform
prior on the variance using the maximum value quoted above.} & \citet{arnaudetal10} & P11-LS \\
$\Mf$--$\Yx$ & 10.0 & $2.25\pm0.12$ & $0.68 \pm 0.04$ & $0.072\pm 0.011$ & \citet{mantzetal10b}  & M10 \\
$\Mf$--$\Yx$ & 4.0 & $1.65$ & 0.6 & --- & Reference & --- \\
\hline
\hline
$D_A^2\Ysz$--$C\Yx$ &	$8.0$ & 	$1.877 \pm 0.028$	& $0.916\pm 0.035$	 & $0.082\pm 0.035$ & \citet{rozoetal12a} & V09 \\ 
$D_A^2\Ysz$--$C\Yx$ &	$10.0$ & $2.341\pm 0.038$ & $0.828\pm 0.057$ 	& $0.167\pm 0.039$ &  \citet{rozoetal12a} & P11-LS(z=0.23) \\ 
$D_A^2\Ysz$--$C\Yx$ &	$10.0$ & $2.100 \pm 0.09$ &	1.0	&  $\leq 0.15$\footnote{Uncertainty in the scatter is implemented as a uniform prior on
the variance with the maximum value quoted above.  The maximum value is chosen to be close to that derived from the \citet{planck11_local}
data by \citet{rozoetal12a}.}  & This work & M10 \\
$D_A^2\Ysz$--$C\Yx$ &	$10.0$ & $2.303$ &	1.0	&  --- & Reference 
\enddata
\label{tab:ref_relations}
\end{deluxetable*}


We consider the cluster scaling relations as measured by three groups.  The first is the P11-LS data set, which employs \planck\ and \xmm\
data, and is comprised
of \citet{planck11_local}, and the associated 
publications based on REXCESS clusters \citep{prattetal09,arnaudetal10}.  We note that in paper I we demonstrated that the $z\leq 0.13$ and
$z\in[0.13,0.3]$ galaxy clusters in \citet{planck11_local} appear to be systematically difference.  Consequently, and motivated by our ultimate goal 
of investigating the maxBCG $\Ysz$--$N_{200}$ scaling relation, we will also be be explicitly 
considering the subset of P11-LS galaxy clusters in the redshift range $z\in[0.13,0.3]$, as appropriate for maxBCG systems.
We refer to this subset of the P11-LS data set as P11-LS(z=0.23).
The second data set, V09, is comprised of
\citet{vikhlininetal09} and \citet{rozoetal12a}, but relies on \chandra\ rather than \xmm\ 
data. Finally, we consider the X-ray analysis of M10 \citep{mantzetal10b}, which is also \chandra\ based.
Because this is the second in a series of papers, we will simply refer the reader to paper I for a more detailed
description of each of the cluster data sets on which the scaling relations analysis is based.  
Here, we focus exclusively on the scaling relations reported in these works.

Table \ref{tab:ref_relations} summarizes the $\Lx$--$M$, $M$--$\Yx$, and $\Ysz$--$\Yx$ scaling relations as quoted in each of the above 
works, though modified when necessary to match the definitions for cluster observables adopted in this work 
(see below).  We always adopt the pivot points reported by each individual work.  The only exception to this rule is that
in the $\Lx$--$M$ relation from V09, where we set $M_0$ to the median mass of the \citet{vikhlininetal09} low redshift cluster sample, 
since the published relation used $1\ \msun$ as its pivot point.  We emphasize that in all cases, given two observables $Y$ and $X$,
our definition of the $Y$--$X$ scaling
relation is the expectation value $\avg{\ln Y|X}$.  So, for instance, in the case of the $\Lx$--$M$ relation derived from X-ray samples,
we only ever consider the Malmquist-bias corrected relations.  
This uniformity of definition is crucial for a self-consistent analysis of multi-variate cluster scaling relations.  When considering
stacked relations, we also take care to correct for the expected difference between $\ln \avg{Y|X}$ and $\avg{\ln Y|X}$ within the
context of a log-normal scatter model.

Because our final goal (paper III) is to use the results from this work to investigate the discrepancy between theory and observations
uncovered by \citet{planck11_local}, 
we have evaluated all scaling relations at $z=0.23$, the median redshift of the maxBCG cluster sample.  The $\Ysz$--$\Yx$
relation in \citet{rozoetal12a} is constrained only at $z \approx 0.1$, and we assume no redshift evolution when extending this result
to $z=0.23$, consistent with self-similar evolution.  
We note that all three works have different evolution terms in their scaling relations.  For instance, for $\Lx$--$M$,
\citet{prattetal09} assume $\Lx \propto E(z)^{7/3}$, while \citet{vikhlininetal09} and \citet{mantzetal10b} 
find a best fit evolution $\Lx \propto E(z)^{1.85\pm 0.42}$ and $\Lx \propto E(z)^{2.34\pm 0.05}$ respectively.  The self-similar expectation 
for soft-band X-ray luminosities is approximately $\Lx\propto E(z)^2$.  The evolution factors of P11-LS and M10 are nearly identical, with
the evolution relative to V09 scaling as $E(z)^{0.49}\approx 1.06$.  That is, the relative evolution from $z=0$ to $z=0.23$
can induce up to a $6\%$ difference between the various works.  
This is also an extreme case, since all scaling relations have a redshift pivot point $z>0$.
Because these $\sim 5\%$ differences are typically much smaller than that the differences between works, we will simply ignore
them here, evaluating all scaling relations at $z=0.23$ using each group's $E(z)$ evolution factors.
Determining what the correct evolution factor is for each scaling relation is beyond the scope of this work.


\begin{deluxetable*}{lccccc}
\tablewidth{0pt}
\tablecaption{Mean Log Differences in X-ray Properties for Sample Pairs}
\tablehead{Property & M10--V09 & P11-LS--V09 & P11-LS--M10  & P11-LS--M10 & P05--V09 \\
& & & Low z & High z }
\startdata
$\Lx$\footnote{Offset computed after outlier removal.} & $0.12\pm 0.02$ & $-0.01\pm 0.02$ & $-0.12\pm 0.02$ & $-0.10\pm 0.03$ & --- \\
$\Mgas$$^a$\footnote{Offset computed after correction to a common aperture.} & $0.03\pm 0.02$ &  $0.03\pm 0.02$ & $-0.02\pm 0.03$ & $-0.04\pm 0.02$ & --- \\
$\Tx$ & --- & $-0.13\pm 0.02$ & --- &  $-0.14\pm 0.05$ & --- \\
$\Yx$$^{ab}$ & --- & $-0.15\pm 0.03$ & --- & $ -0.19 \pm 0.05$ & --- \\
$\Mf$\footnote{Relaxed/cool core only.}\footnote{Relative to paper I, the offsets here are corrected by an $11\%$
shift due to an update in \chandra\ calibration.  This update was not applied to the masses quoted in the Tables in M10, but is applied for the scaling relations.  
} 
	& $-0.03 \pm 0.02$ &  $-0.12\pm 0.02$ & $-0.05\pm 0.07$ & $-0.37\pm 0.07$ & $-0.18\pm 0.05$ \\
$\Mf$$^d$\footnote{Non-relaxed/no cool core only.}  & $0.11 \pm 0.11$ &  $-0.14\pm0.03$  & $-0.14\pm 0.03$  & $-0.34\pm0.06$ & ---
\enddata
\label{tab:offsets}
\end{deluxetable*}


We have also homogenized
all data to common definitions:  $\Lx$ is defined as the total X-ray luminosity within $\Rf$ in the $[0.1,2.4]\ \keV$ band, $\Tx$
is defined as the spectroscopic temperature in an an annulus $R\in[0.15,1]\Rf$, $\Mgas$ is the total gas mass within $\Rf$,
$\Yx = \Mgas\times \Tx$, and $\Ysz$ is the integrated Compton parameter within a sphere of radius $\Rf$.
Whenever our definitions do not match the definitions employed in any of the works we consider, all data is rescaled as described in paper I.
The data in Table \ref{tab:ref_relations} --- in conjunction with the systematic offsets we estimated in paper I --- 
form the basis of our analysis.

The different groups rely on different analysis techniques and different mass proxies when estimating cluster
scaling relations.  For the $\Lx$--$M$ analysis, V09 and P11-LS \citep[via][]{prattetal09} rely on $\Yx$ as a mass proxy, while Mantz
relies on $\Mgas$.   In all cases, the mass-proxy ($\Yx$ or $\Mgas$) relation with mass is calibrated using hydrostatic
mass estimates of relaxed galaxy clusters.
V09 employs a likelihood fitting method that explicitly incorporates
Malmquist bias, and tests their fitting routine using Monte Carlo cluster samples.  The M10 analysis is very
similar in spirit to the V09 analysis, but also explicitly incorporates the cosmological dependence of the cluster abundance function in the
cluster scaling relations.
Finally, \citet{prattetal09} quote two different fits, corresponding to $BCES(Y|X)$ and $BCES$-orthogonal fits.  $BCES(Y|X)$
is closest to a likelihood fit in the absence of Malmquist bias.  Since this fit is the one that is most comparable to 
those of V09 and M10, we only employ the $BCES(Y|X)$ Malmquist-bias corrected fit of \citet{prattetal09} in this 
analysis.  We caution that this is still not directly comparable to likelihood based approaches, since neglecting
intrinsic scatter in the cost function being minimized gives clusters with small errors undue weight in the 
fit \citep[see also][]{andreonhurn10}.\footnote{We see very clear evidence of this, for instance,
when measuring the $\Ysz$--$\Lx$ relation from the $z\in[0.12,0.3]$ clusters in \citet{planck11_local}: a $BCES(Y|X)$ fit results
in steeper scaling relations than a likelihood fit because a few very massive clusters with extremely tight error bars dominate the penalty
function being minimized.  By including intrinsic scatter as a free parameter in the fit --- as opposed to computing it a posteriori
based on a fit including only statistical errors --- these clusters are de-weighted, which flattens
the resulting relations.}  Importantly, as we show below, these differences in fitting methodology are not the primary drivers
of the tension between the various scaling relation analyses.

Turning to the $M$--$\Yx$ relation, \citet{arnaudetal10} rely on hydrostatic mass estimates of relaxed galaxy clusters
to calibrate $M$--$\Yx$.  Because \citet{arnaudetal10} did not quote the corresponding scatter, we rely instead on the
value quoted in their previous work, \citet{arnaudetal07}.  
\citet{vikhlininetal09} followed a similar approach to determine $M$--$\Yx$, but also fail to quote a constraint on the
scatter, noting only that the data is consistent with no scatter, as well as with the predictions from \citet{kravtsovetal06}.
Thus, we place an upper limit on the scatter based on the value reported in \citet{kravtsovetal06}.
\citet{mantzetal10b} uses $\Mgas$
as a mass proxy, and the $M$--$\Mgas$ relation is itself calibrated based on hydrostatic mass estimates
of relaxed galaxy clusters.  We note too
that \citet{mantzetal10b} report $\Yx$--$M$ rather than $M$--$\Yx$.  To obtain $M$--$\Yx$,
we simply invert their relation.  The corrections due to scatter are $2\%$ in the amplitude (see equation \ref{eq:avgms}),
and are completely negligible both
relative to their quoted uncertainty, and for the purposes of this work.

Finally, turning to the $\Ysz$--$\Yx$ scaling relation, we write this as
\be
\avg{\ln \left(D_A^2\Ysz\right) |C\Yx} = a_{SZ} +\alpha_{SZ} \ln C\Yx.
\ee
where $D_A$ is the angular diameter distance, and $C$ is a normalization constant,
\be
C = \frac{\sigma_T}{m_ec^2} \frac{1}{\mu m_p} = 1.407\times \frac{10^{-5}\ \Mpc^2}{10^{14}\ \msun\keV}, \label{eq:c}
\ee
where $\sigma_T$ is the Thompson scattering cross section, $m_e$ is the mass of the election, $m_p$ is the
mass of a proton, and $\mu=1.156$ is the mean molecular weight of the intra-cluster medium for solar abundance
(variation in metallicities indues a $\approx 1\%$ uncertainty, completely negligible for our purposes).

Our constraints for the $\Ysz$--$\Yx$ scaling relation come from two sources. \citet{rozoetal12a}
constrain the $\Ysz$--$\Yx$ ratio using \planck\ and \chandra\, and this is the result we use
later to estimate the $\Ysz$--$M$ relation in conjunction with the \citet{vikhlininetal09} data.
Turning to the P11-LS data set, we noted in paper I that the P11-LS $\Ysz/\Yx$ ratio is different for
$z\leq 0.13$ and $z\ge 0.13$ clusters, so rather than using the fit from \citet{planck11_local} for the
$\Ysz$--$\Yx$ relation, we have refit the P11-LS data for galaxy clusters in the $z\in[0.13,0.3]$
redshift range using the method of \citet{rozoetal12a}.  Our best fit scaling relation for the P11-LS(z=0.23)
data is included in Table \ref{tab:ref_relations}.

It is important to emphasize that by choosing to use the $\Ysz$--$\Yx$ scaling relation derived from the P11-LS
data in our predictions for the $\Ysz$--$M$ (and eventually $N_{200}$--$M$) scaling relations,
we are diverging from the treatment of \citet{planck11_optical} for deriving the $\Ysz$--$\Nt$ scaling relation.
In that work, the $\Ysz$--$M$
relation used was that of \citet{arnaudetal10}, which was derived from X-ray clusters alone.
A more detailed discussion about the X-ray expectations for the $\Ysz$--$M$ scaling relation
is presented in \citet{rozoetal12a}.  Here, the point that we want to emphasize is that by working
with the {\it observed} $\Ysz$--$\Yx$ scaling relation, we are explicitly testing whether all scaling
relations are self-consistent, in the sense that a multivariate log-normal model can explain 
all observational data.  Discussion of the consistency of the $\Ysz$--$Y_X$ scaling relation with 
X-ray expectations can be found in \citet{rozoetal12a}.

For completeness, we have also estimated the $D_A^2\Ysz/C\Yx$ ratio that would be derived from a combination of the \citet{planck11_local}
data and the \citet{mantzetal10b} data (i.e. we are explicitly setting $\alpha_{SZ}=1$ for the M10 data set).
For all galaxy clusters shared by the M10 and P11-LS cluster samples, using the formalism from paper I 
we rescale the $M_{gas}$ values in M10 to the $R_{500}$ aperture adopted by P11-LS.  
We then directly compute $Y_X$ for these systems, and use bootstrap resampling 
to compute the mean value of $\ln(D_A^2\Ysz/C\Yx)$ at this aperture, along with the associated uncertainty.
We find $\avg{\ln(\Ysz/\Yx)} = -0.16\pm 0.09$.  
We have confirmed that applying this exact method to the \citet{vikhlininetal09} data results in $\ln(\Ysz/\Yx)=-0.20 \pm 0.04$,
in perfect agreement with the more detailed analysis of \citet{rozoetal12a}.  We do not provide a scatter estimate since
the errors in \citet{mantzetal10b} represent the expected uncertainty in $\Yx$, rather than just measurement errors.  E.g.
the error bar includes the expected scatter in gas clumping from cluster to cluster, which alone contributes $\approx 5\%$ scatter,
already larger than the typical measurement error.  
The upper limit to the scatter quoted in Table \ref{tab:ref_relations} is {\it assumed} based on the
scatter derived from the \citet{planck11_local} data by \citet{rozoetal12a}.  Note, however, that 
comparison to the P11-LS and V09 values suggests the quoted upper limit is
reasonable.
The best fit $D_A^2\Ysz/C\Yx$ is $0.82$, corresponding to an amplitude parameter $a=2.10$ for the $\Ysz$--$\Yx$
relation.

For future reference, Table \ref{tab:offsets} shows the offsets in X-ray derived cluster observables from paper I.
In fact, table \ref{tab:offsets} is a reproduction of Table 3 from paper I, albeit with some slight modifications.
First, because the $\Lx$--$M$ relation of \citet{prattetal09}
relies on the mass calibration of \citet{pointecouteauetal05} --- henceforth P05 --- we have also explicitly added the mass offset between
P05 and V09 for the purposes of our comparison (the offset was computed in paper I, but not included in Table 3 of that work).  
In addition, the scaling relations in \citet{mantzetal10b}
were corrected for a \chandra\ calibration update, which uniformly shifts the masses of galaxy clusters by $\approx 11\%$.
Because the cluster data tables in \citet{mantzetal10b} were {\it not} corrected for this calibration
update, but the scaling relations were, we have corrected the observed P11-LS--M10 and M10--V09 offsets from paper I values to reflect the calibration employed by \citet{mantzetal10b} in their scaling relation analysis.  This calibration update significantly improves the agreement between the P11-LS and M10 masses.  
Because Mantz quotes \rosat-calibrated luminosities and $\Mgas$, 
there is no corresponding shift in $\Lx$ or $\Mgas$.   While $\Mgas(\Rf)$ does shift due to aperture
effects, Table \ref{tab:offsets} quotes the $\Mgas$ offset at a fixed aperture, so an aperture correction is not necessary.


\section{Comparison of Cluster Scaling Relations}
\label{sec:cluster_relations}

Below, given a scaling relation $Y$--$X$, we will compare the $68\%$ confidence regions in the $Y$--$X$ plane 
corresponding to the probability distribution $P(\ln Y|\ln X)$.  These regions are estimated by performing
$10^5$ Monte Carlo draws of the amplitude and slope of each published scaling relation.  We then
evaluate $\avg{\ln Y|\ln X}$ along a logarithmically spaced grid in $X$, and compute the standard deviation
of the resulting $\avg{\ln Y|\ln X}$ values at each point.  This defines our estimate of the 68\% confidence region
of each cluster scaling relation in the $Y$--$X$ plane.

\subsection{The $\Lx$--$M$ Relation}
\label{sec:ref_lxm}

\begin{figure}
\begin{center}
\scalebox{1.2}{\plotone{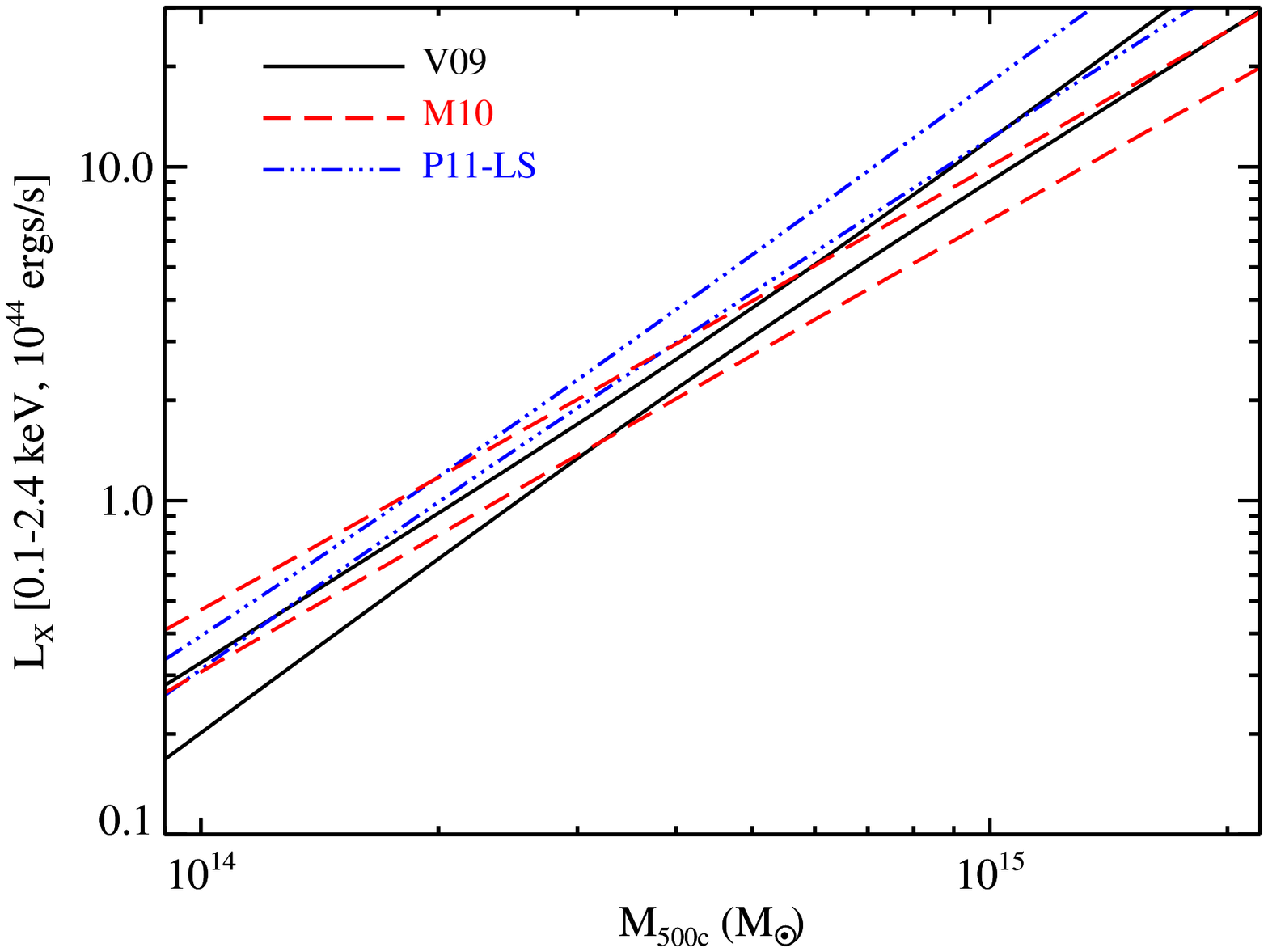}}
\scalebox{1.2}{\plotone{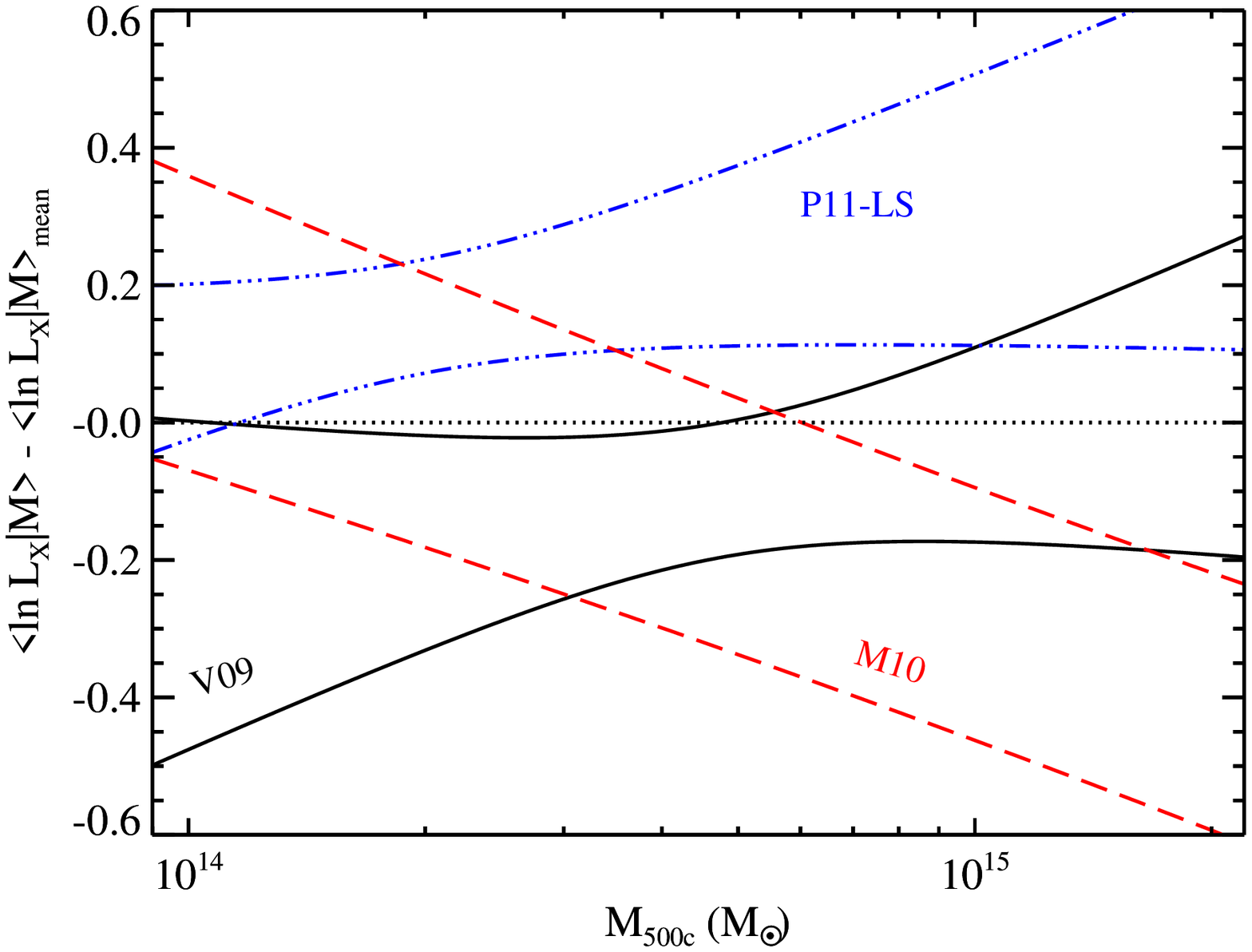}}
\caption{{\it Top panel: }$68\%$ confidence band for the $\Lx$--$M$ relation for the
P11-LS, V09, and M10, data sets, as labelled.
All scaling relations are evaluated at the median
redshift of the maxBCG sample, $z=0.23$.  
{\it Bottom panel:} As per the top-panel, after subtracting out a reference mean scaling
relation (see text for the definition of the mean scaling relation).
Tensions are evident among the scaling relations.
}
\label{fig:lxm}
\end{center}
\end{figure}

The top panel in Figure \ref{fig:lxm} compares the $\Lx$--$\Mf$ relations
derived from the P11-LS, V09, and M10 data sets.
We remind the reader that the $L_X$--$M$ relation we associate with the P11-LS data set
is that of \citet{prattetal09}.
The width of these bands reflects the
quoted errors in each of the works being compared, as summarized in Table \ref{tab:ref_relations}.  

To more easily compare these relations, we compute the 
offset of each relative to a mean relation.  The amplitude of the mean relation is defined 
as the average value of $\ln \Lx$, evaluated at mass $\Mf = 4\times 10^{14}\ \msun$, as estimated from 
the P11-LS, V09, and M10 best fit scalings.  
The slope of the mean relation is likewise defined as the mean of the slopes
of the three scaling relations.  This mean scaling relation is only intended to be a reference to be subtracted from the three scaling relations:
we do not ascribe any physical interpretation to it. 
More specifically, we do not advocate interpreting the mean scaling as more likely to be correct than any of its constituent elements.  

The M10 errors of Table \ref{tab:ref_relations} are larger in amplitude than those quoted for the P11-LS
and V09 data sets because they include systematic uncertainties, whereas the V09 and P11-LS errors are measurement errors only.   
The published M10 slope, however, has an error of only $0.05$, compared with $0.11$ and $0.14$ for P11-LS and V09, respectively.   
These differences in the relative errors of the amplitude and slope explains why the P11-LS and V09 contours have 
an hourglass shape, while the M10 contours are nearly parallel lines.


The lower panel in Figure \ref{fig:lxm} shows the $68\%$ confidence intervals of the $\Lx$--$M$ relations after
subtracting the mean relation.
We see that the P11-LS scaling relation is parallel to but
clearly offset from the V09 scaling relation, while the M10 relation is  flatter. 
We can test whether the difference in amplitude between these various scaling relations are due to
differences in the raw X-ray cluster observables simply by shifting the amplitude of all the
relations by the systematic offsets quoted in Table \ref{tab:offsets}.  
If  $\Delta \ln \Lx = \avg{\ln \Lx}_{\rm B} - \avg{\ln \Lx}_{\rm A}$ is the mean offset between samples 
$B$ and $A$ in log-luminosity, and with $\Delta \ln M$ defined similarly for total mass, then the scaling relation 
from sample $A$ can be observable-offset corrected to sample B by adding the constant $[\Delta \ln \Lx -\alpha_A \Delta \ln M]$ to $\avg{\ln \Lx}_{\rm A}$.  
Assuming that the only difference between the two data sets $A$ and $B$ are the observable offsets characterized
in paper I, applying these corrections to every data set should bring all scaling relations into good agreement with each other.
For specificity,
we observable--offset correct every scaling relation to the P11-LS scaling relation.

These amplitude shifts do not affect the slopes of the relations, and therefore
cannot explain the tension between the slope derived by M10 and
the remaining two data sets.   Instead, the difference in slope
can be traced to differences in $\fgas$--$M$ between the
various works.  Based on the results by \citet{allenetal08}, \citet{mantzetal10b}
assume that $\fgas = f_0$, a constant value.  In paper I, we saw that given a different, 
power-law model, $\fgas(M) \sim M^\gamma$, the relation between the mass $M$ of that model and that estimated by M10 is
\be
M_{M10} = M \left( \frac{f_{\rm gas}(M)}{f_0} \right)^{1.67} = M \left( \frac{M}{M_p} \right)^{1.67\gamma}
\ee
where $M_p$ is the pivot mass of the $\fgas$--$M$ relation, and $\gamma$ is the index of the $\fgas$--$M$ power-law.
If $\psi$ is a cluster observable where the M10 scaling relation is $\psi \propto M^{\alpha_{M10}}$,
then the scaling relation the M10 scaling relation for an $\fgas\propto M^\gamma$
would be
\be
\psi=A\left[ \left(\frac{M}{M_p}\right)^{1.67\gamma} \frac{M}{M_p} \right]^{\alpha_{M10}}.
\ee
Here, $M_0$, $A$, and $\alpha_{M10}$ are the pivot point, amplitude, and slope of the $\psi$--$M$ relation reported by \citet{mantzetal10b}.
The slope $\alpha$ relative to the mass $M$ in this new $\fgas$ model is 
\be
\alpha = \alpha_{M10}(1+1.67\gamma).
\ee
In the work of V09 and P11-LS, the slope $\gamma$ of the $\fgas$--$M$ relation is $\gamma\approx 0.1-0.2$. Setting $\gamma=0.15$,
and given $\alpha_{M10}=1.34\pm 0.05$,
our model predicts an $\Lx$--$M$ slope $\alpha=1.68\pm 0.05$, resulting in an excellent match to the values of P11-LS and V09.
Evidently, the reason that the M10 scaling relation appears flatter is that the $\fgas$ model employed by M10 is not consistent
with that of P11-LS or V09.


\begin{figure}
\begin{center}
\scalebox{1.2}{\plotone{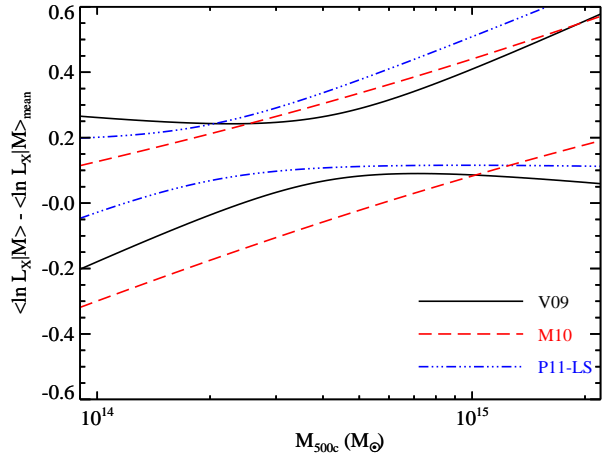}}
\caption{$68\%$ confidence band for the $\Lx$--$M$ relation derived
from the P11-LS, M10, and V09 data sets, as labelled.  
We have subtracted out a reference scaling relation as per the bottom
panel of Figure \ref{fig:lxm}.
All scaling
relations have been displaced along the $\Lx$ and $M$ axis by the 
observable offsets relative to the P11-LS and P05 data sets
characterized in Paper I.   In addition, the
M10 scaling relation has been tilted from a constant $\fgas$ model
to a model with $\fgas \propto M^{0.15}$.  These corrections improve 
agreement among the scaling relations.  
}
\label{fig:lxm_shifted}
\end{center}
\end{figure}


Figure \ref{fig:lxm_shifted} shows the difference in the $\Lx$--$M$ relation of V09, M10, and P11-LS,
after accounting for the observable offsets  in Table \ref{tab:offsets}.  The M10 relation is also tilted
to an $\fgas\propto M^{0.15}$ model as described above.\footnote{We note that when applying a shift
in the mass-axis and a tilt, it matters whether we shift first or tilt first.  A uniform mass shift would only
be appropriate for parallel relations, so we tilt first, then shift. }  
For M10, we show the $\Lx$--$M$ relation after shifting
the M10 values using the P11-LS--M10 offset (dashed lines), and then also applying the offset between P11-LS and P05,
which is necessary because the \citet{prattetal09} scaling relation is calibrated based on P05.  
With these corrections, all scaling relations are in excellent agreement with each other.
We conclude that the differences between the three works
are driven by  systematic offsets in X-ray observables, and the slope of the $\fgas$--$M$ relation.
In particular, the statistical treatment of each of the samples --- i.e. the fact that each data
set used a different statistical method for fitting the scaling relations and correcting for selection effects ---
appears to play a secondary role relative to the overall level of systematic offsets in the X-ray observables themselves.


\subsection{The $M$--$\Yx$ Relation}
\label{sec:ref_yxm}


\begin{figure}
\begin{center}
\scalebox{1.2}{\plotone{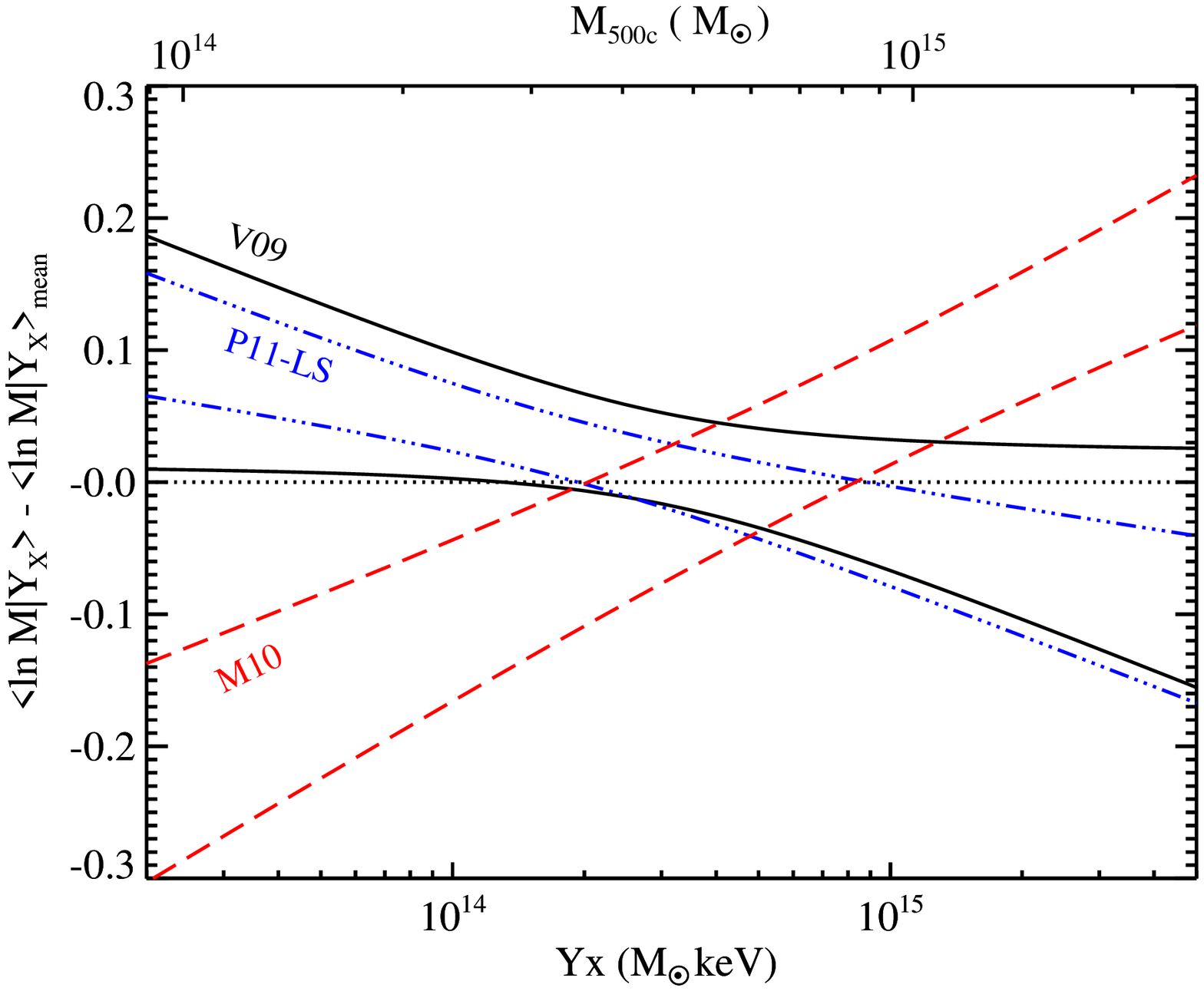}}
\scalebox{1.2}{\plotone{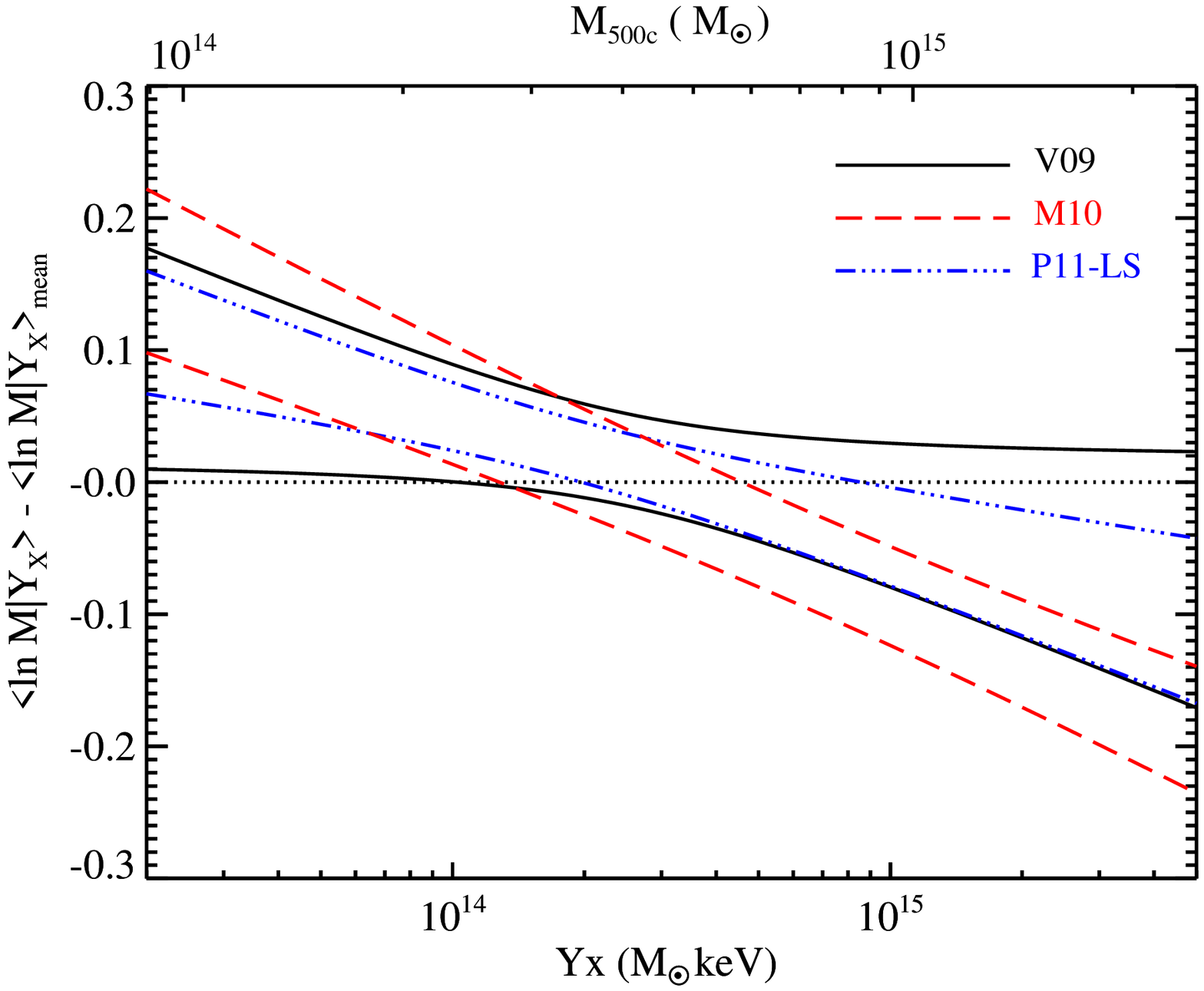}}
\caption{{\it Top panel: } $68\%$ confidence band for the $M$--$\Yx$ relation derived
from the P11-LS, M10, and V09 data sets.
For ease of comparison, we subtract out a reference mean relation as per the bottom panel in Figure \ref{fig:lxm}.
{\it Bottom panel: } As top panel, but after correcting for the systematic differences in cluster observables,
and tilting the Mantz relation to that of an $\fgas\propto M^{0.15}$ model.  As in the case of the
$\Lx$--$M$ relation, all scaling relations are now in good agreement.
}
\label{fig:myx}
\end{center}
\end{figure}


The top panel in Figure \ref{fig:myx} shows the $68\%$ confidence region for the
$M$--$\Yx$ relation of the three works considered here, after subtracting
a mean scaling relation defined by a self-similar slope of $3/5$, and setting the amplitude to the mean
amplitude of three works considered here at $\Yx=4\times 10^{14}\ \msun\keV$.  
The parameters for this mean scaling relation are given in Table \ref{tab:ref_relations}. 
The V09 and P11-LS relations are in excellent agreement,
despite the large differences in the cluster-by-cluster comparison of mass and $\Yx$ measurements, demonstrating
that the systematic offsets from paper I move clusters along this best-fit relation.   Unfortunately, because we do not know what
is the source of the P11-LS--V09 offset in cluster observables, it is difficult to guess why this ``conspiracy of errors'' would happen.
As for the M10 relation, we once again see a significantly different slope, so the amplitude offset depends on the
mass scale under consideration.

The P11-LS--V09 comparison argues against instrumental calibration being the principal
source of the observed systematic errors.  Specifically, a temperature bias, $b_T$, induces a bias $b_M=b_T^{1.5}$ in hydrostatic masses \citep{vikhlininetal06}.   This in turn induces a bias $b_{\rm gas} = b_M^{0.4} = b_T^{0.6}$ via aperture effects, so the net bias in $\Yx$ is $b_Y=b_T^{1.6}$.  Letting $\alpha\approx 5/3$ be the slope of the $\Yx$--$M$
relation, we see that a temperature calibration bias $b_T$ will induce a bias $b_T^{1.5\alpha-1.6} \approx b_T^{0.9}$
in the amplitude of the $\Yx$--$M$ relation.  In other words, an overall temperature bias due to instrumental calibration
does not appear to shift galaxy clusters along the scaling relation.

The lower panel in Figure \ref{fig:myx} illustrates the offset-corrected scaling relations.  We have also tilted the M10
relation to that of a $\gamma=0.15$
$\fgas$ model.  As in the case of  $\Lx$--$M$, after these corrections all three scaling relations are in
agreement within the expected errors.


\subsection{The $\Ysz$--$\Yx$ Scaling Relation}
\label{sec:ysz_prediction}


\begin{figure}
\begin{center}
\scalebox{1.2}{\plotone{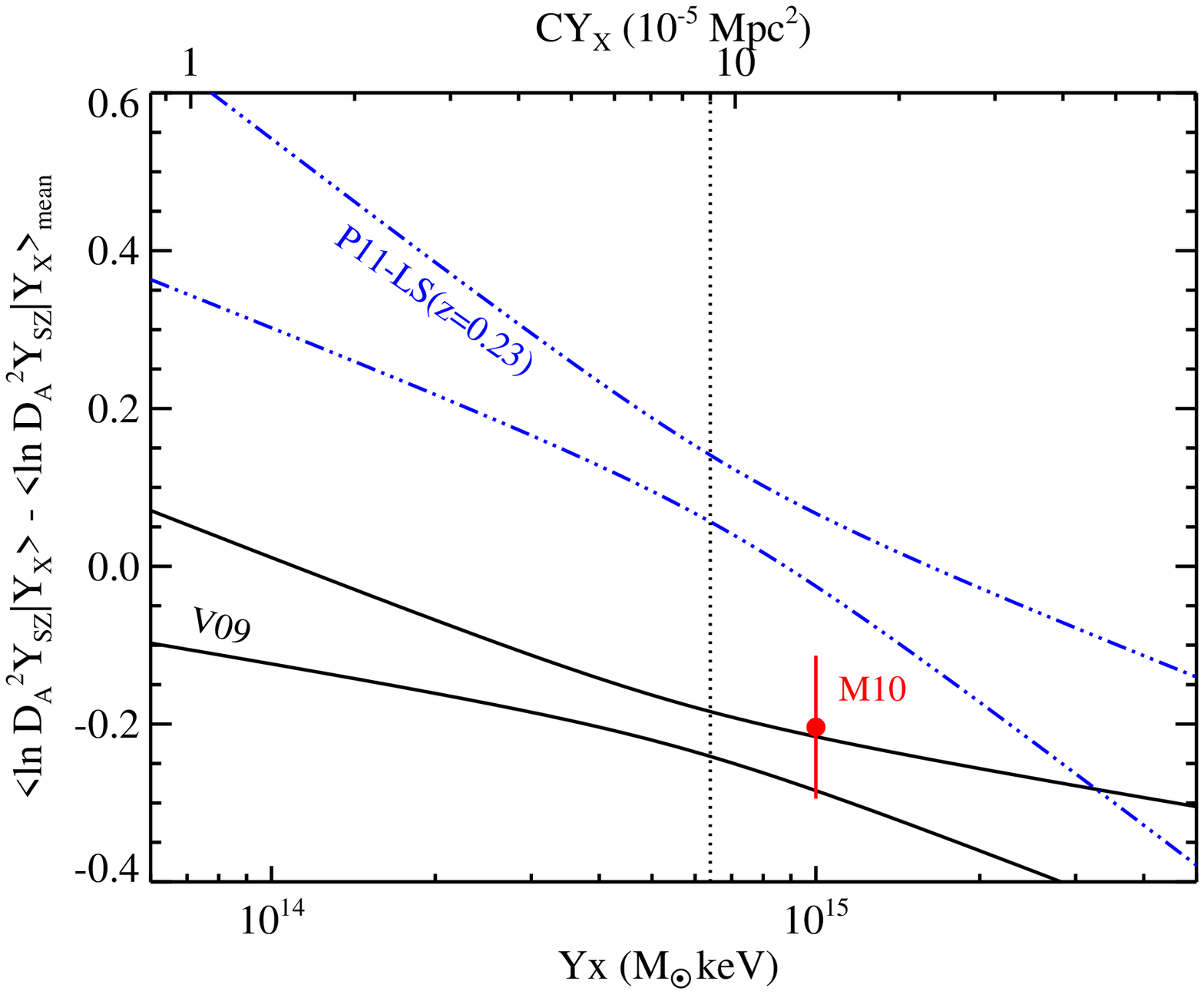}}
\scalebox{1.2}{\plotone{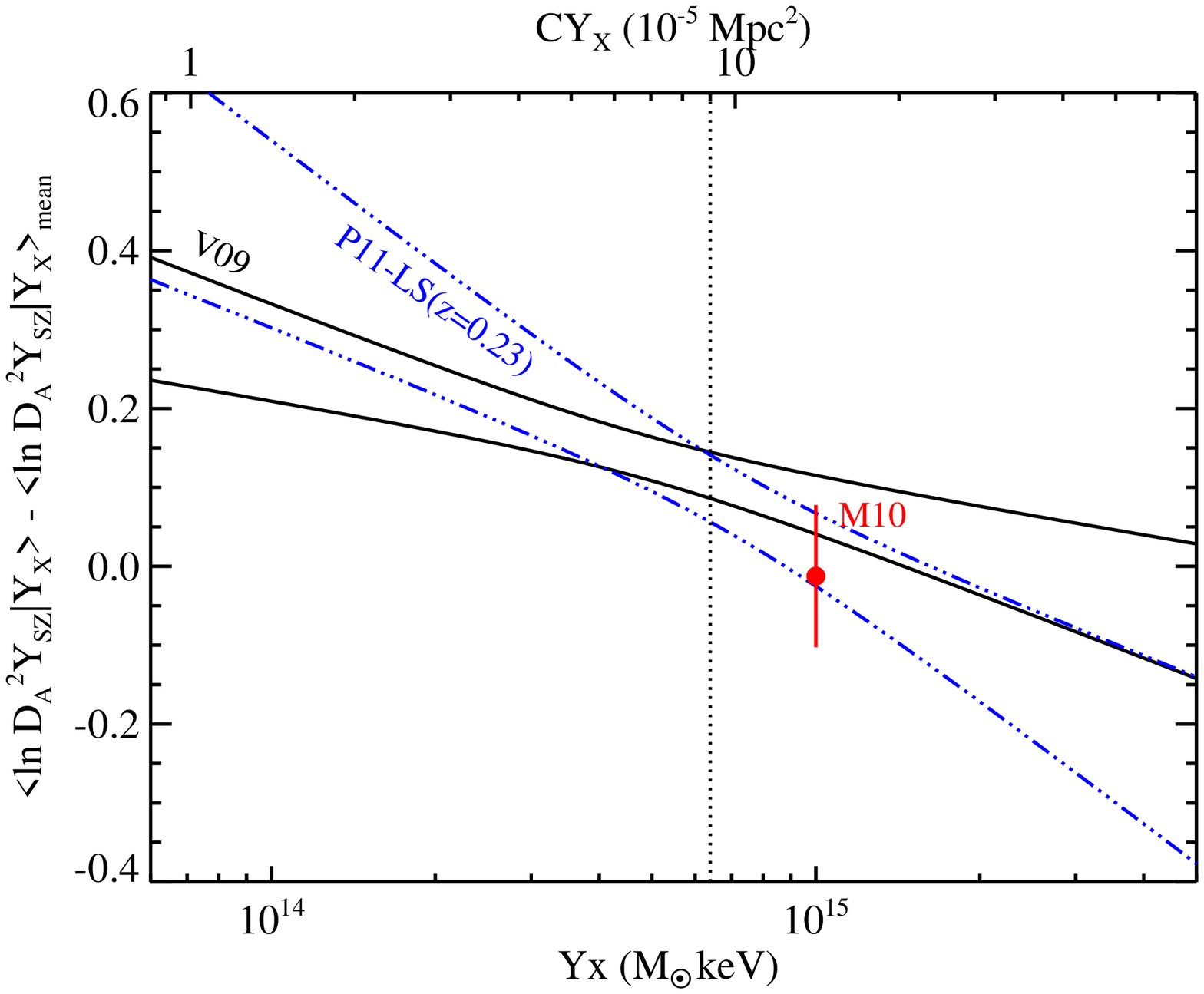}}
\caption{
{\it Top panel: } The $\Ysz$--$\Yx$ relation at $z=0.23$ for each of the data sets we consider, as labelled.
There are large differences between each of the data sets.
{\it Bottom panel: } The $\Ysz$--$\Yx$ relation after correcting for cluster observable systematic offsets, and
correcting the Vikhlinin curve for the expected redshift evolution.  The vertical dotted line is the pivot point
of the V09 relation. 
}
\label{fig:yszyx}
\end{center}
\end{figure}


The top panel in Figure \ref{fig:yszyx} shows the $\Ysz$--$\Yx$ scaling relation after subtracting out the reference relation
$\avg{\ln D_A^2\Ysz|\Yx} = \ln(C\Yx)$ where $C$ is given by equation \ref{eq:c}, and assuming both $D_A^2\Ysz$ and $C\Yx$
are measured in the same units.  
Because we assume $\alpha=1$ for the Mantz data, we only place a single point with error bars at the 
pivot point of the sample.  There are large differences between the P11-LS(z=0.23) and the V09 and M10 samples.

The bottom panel in Figure \ref{fig:yszyx} corrects the V09 and M10 scaling relations for the observable offsets from
Table \ref{tab:offsets} relative to the P11-LS(z=0.23) data.
The V09 scaling relation is modified using the shifts relative to the low-redshift P11-LS data, and then again by the redshift correction
between low and high redshift based on the $\Ysz/\Yx$ ratio.  
The M10 value is shifted using the $\Yx$ shift relative to P11-LS(z=0.23) directly.
With these corrections, all data sets agree on the amplitude of the $\Ysz/\Yx$ relation, though some differences in the slope
are apparent. 

\subsection{Summary of Results}

The main result of this section is very clear: despite differences in fitting methods as well as differences in the treatment of selection
effects, after accounting for the systematic differences in derived X-ray cluster observables and modeling of the slope of the $M_{gas}$--$M$ relation, 
all scaling relations are in good agreement
with each other.  In other words, systematics in the treatment of
selection effects and/or fitting methods are clearly sub-dominant to the overall systematic offset in the input data used to
derive the scaling relations.


\section{The $\Ysz$--$M$ Relation as Calibrated from X-rays}
\label{sec:yszm}

Having specified the $M$--$\Yx$ and $\Ysz$--$\Yx$ relation, the $\Ysz$--$M$ relation
is nearly completely specified.  We now derive expressions for the amplitude and slope
of the $\Ysz$--$M$ scaling relation in terms of the others, which we then apply to each of the X-ray data sets considered in this work.    

Part of the motivation
for this work is simply to transport the $M$--$\Yx$ calibration from X-ray data sets
to $\Ysz$--$M$, in order to enable cosmological interpretations of SZ cluster samples.
Just as importantly, however, we wish to determine the differences in the predicted
$\Ysz$--$M$ scaling relation for each of these data sets, as these differences
will necessarily have an important impact in the interpretation of the SZ--richness
scaling relations of maxBCG galaxy clusters (paper III).  

\subsection{Method}

We define
\bea
y_x & = & \ln \left( \Yx/Y_{0} \right) \\
\ysz & = & \ln \left( D_A^2\Ysz/\Delta_0^2\right) \\
m & = & \ln \left(M/M_0\right) \\
\eea
where the normalization constants $Y_0$, $\Delta_0^2$, and $M_0$ are chosen so as to
decorrelate the amplitude and slope of the cluster scaling relations.  
In the previous section, we specified the $\Ysz$--$\Yx$ and $M$--$\Yx$ scaling relation.
Combining this two relations, we can arrive at an expression for the $\Ysz$--$M$ scaling
relation.  Specifically, let us write
\bea
\avg{\ysz|m} & = & a_{sz|m} + \alpha_{sz|m}m  \label{eq:yszm}\\
\avg{m|y_x} & = & a_{m|x} + \alpha_{m|x}y_x \label{eq:myx} \\
\avg{\ysz|y_x} & = & a_{sz|x} + \alpha_{sz|x}y_x \label{eq:yszyx} 
\eea
We wish to solve for $a_{sz|m}$ and $\alpha_{sz|m}$ in terms of known parameters.
Using a local power-law model for cluster abundances --- see Appendix \ref{app:multi},
and specifically equations \ref{eq:avgms} and \ref{eq:avgs2s1} \citep[see also][]{whiteetal10} --- one
has that the parameters of the various relations above are related via
\bea
a_{sz|x} &= & \left[ a_{sz|m}+\alpha_{sz|m}a_{m|x}\right] \nn \\
	& & \hspace{0.2 in} + r_{sz,x|m}\beta \alpha_{sz|m}\sigma_{m|x}\sigma_{m|sz} \label{eq:szamp}\\
\alpha_{sz|x} & = & \alpha_{sz|m}\alpha_{m|x}
\eea
where $r_{sz,x|m}$ is the correlation coefficient between $\ysz$ and $y_x$ at fixed $m$, $\beta$
is the slope of the halo mass function ($dn/d\ln M \propto M^{-\beta}$), and $\sigma_{m|x}$
and $\sigma_{m|sz}$ are the scatter in $m$ at fixed $y_x$ and $\ysz$ respectively.
Note the term in square brackets in equation \ref{eq:szamp} is simply the naive relation
obtained from plugging equation \ref{eq:myx} into equation \ref{eq:yszm}.

Since both $\alpha_{m|x}$ and $\alpha_{sz|x}$ are known from the previous section, we can readily
solve for $\alpha_{sz|m}$,
\be
\alpha_{sz|m} = \frac{\alpha_{sz|x}}{\alpha_{m|x}}. \label{eq:szm_slope}
\ee
To solve for $a_{sz|m}$ from equation \ref{eq:szamp}, we must first be able to estimate the correction
term depending on the correlation coefficient, which requires that we know both $r$ and $\sigma_{m|sz}$.
While equation \ref{eq:scatrel} in Appendix \ref{app:multi} relates these two quantities, one still needs an
additional independent input to close the system.

We arrive at an additional relation between $r$ and $\sigma_{m|sz}$ using a simple model for how $\Ysz$ and $\Yx$
are correlated.  Motivated by the fact that \citet{rozoetal12a} find that the intrinsic scatter in the $\Ysz$--$\Yx$
relation is consistent with zero,
we make the assumption that the noise in $\ysz$ has two contributions: an intrinsic contribution $\delta y_{int}$,
which is identical and perfectly correlated with the intrinsic noise in $y_x$, reflecting variations in the details
of the physical properties of the intra-cluster medium, and a an additional noise
$\delta y_{lss}$ that is due to structures along the line of sight to the clusters, and which is uncorrelated with fluctuations in $y_x$.  
Setting $\avg{\delta y_{int}^2}=\sigma_{x|m}^2$,
$\avg{\delta y_{int}\delta y_{lss}}=0$, $\delta \ysz = \delta y_{int} + \delta y_{lss}$, and $\delta y_x = \delta y_{int}$, we find
\be
r_{sz,x|m} = \frac{ \avg{\delta \ysz\delta y_x} }{\sigma_{x|m}\sigma_{sz|m}} = \frac{\sigma_{x|m}}{\sigma_{sz|m}}.
\label{eq:rrel}
\ee
In addition, from equation \ref{eq:scatrel_1d} in Appendix \ref{app:multi}, we have 
\be
\sigma_{m|x} = \alpha_{m|x}\sigma_{x|m},
\label{eq:scat1}
\ee
with a similar relation for $\ysz$.   Putting these two together, we can rewrite equation \ref{eq:szamp}
as
\be
a_{sz|m} = \left[a_{sz|x} - \alpha_{sz|m}a_{m|x} \right] - \frac{\beta}{\alpha_{m|x}}\sigma_{m|x}^2.  \label{eq:szamp_final}
\ee
As a rough order of magnitude, we expect a slope of the halo mass function $\beta\approx 3$, and
$\alpha_{m|x}\approx 3/5$, which for $10\%$ scatter in mass corresponds to a nearly negligible 5\% amplitude
shift.
Together, equations \ref{eq:szamp_final} and \ref{eq:szm_slope} completely specify the
$\Ysz$--$M$ scaling relation in terms of known parameters (i.e. in terms of the $\Ysz$--$\Yx$ and
$M$--$\Yx$ scaling relation parameters).

\begin{figure}
\begin{center}
\hspace{-0.2in} \scalebox{1.2}{\plotone{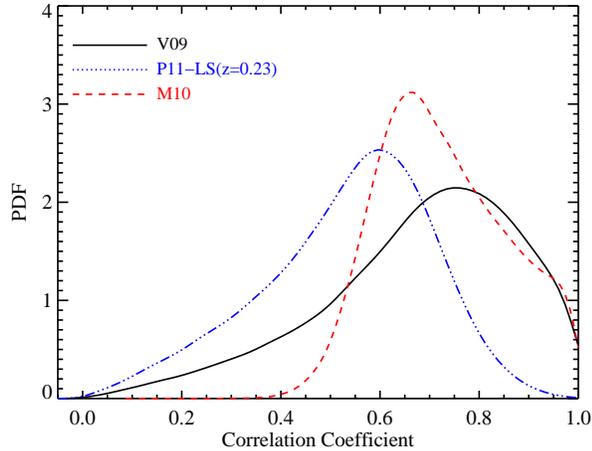}}
\caption{
Likelihood distribution for the correlation coefficient $r\equiv \Cov(\ysz,y_x)/\sigma_{x|m}\sigma_{sz|m}$ for each of the
three data sets we consider, as labelled.  The maximum likelihood value for $r$, as well as the upper and lower limits
defining the $68\%$ confidence contour, are summarized in Table \ref{tab:derived_relations}.
}
\label{fig:r}
\end{center}
\end{figure}

Let us now turn to the scatter in the $\Ysz$--$M$ scaling relation.  From equation \ref{eq:scatrel}
in Appendix \ref{app:multi} we have 
\be
\sigma_{sz|x}^2 = \alpha_{sz|m}^2\left[ \sigma_{m|x}^2 + \sigma_{m|sz}^2 - 2r_{sz,x|m}\sigma_{m|x}\sigma_{m|sz} \right]. \label{eq:scateq}
\ee
Using equations \ref{eq:rrel} and \ref{eq:scatrel_1d}, we can rewrite the $r$ term above in terms of $\sigma_{m|x}$ only.
Solving for $\sigma_{m|sz}^2$, and using equation \ref{eq:scatrel_1d} again to obtain $\sigma_{sz|m}$ from $\sigma_{m|sz}$
we arrive at
\be
\sigma_{sz|m}^2 = \sigma_{sz|x}^2 - \sigma_{m|x}^2 \frac{\alpha_{sz|x}(\alpha_{sz|x}-2)}{\alpha_{m|x}^2}. \label{eq:szm_scat} 
\ee
This gives $\sigma_{sz|m}$ in terms of known quantities.  To solve for $r_{sz,x|m}$, we again use equations \ref{eq:rrel} and \ref{eq:scatrel_1d} to 
write $\sigma_{sz|m}$ in terms of $r_{sz,x|m}$ and $\sigma_{m|x}$, and solve for $r_{sz,x|m}$.  We arrive at
\be
r_{sz,x|m} = \left[ \alpha_{m|x}^2 \frac{\sigma_{sz|x}^2}{\sigma_{m|x}^2} + \alpha_{sz|x} \left(2-\alpha_{sz|x}\right) \right]^{-1/2}.
\ee
We evaluate the right hand side of the above equation using the central values from the V09 
data set to arrive at
$r_{sz,x|m}=0.84$.  As expected, the fluctuations about the mean in $\Ysz$ and $\Yx$ at fixed mass are heavily correlated.
The likelihood for the correlation coefficient $r$ is shown in Figure \ref{fig:r}, and
the corresponding constraint is summarized in Table \ref{tab:derived_relations}.


Having derived analytical relations between the input $M$--$\Yx$ and $\Ysz$--$\Yx$ scaling
relation parameters, and the $\Ysz$--$M$ scaling relation parameters,
 we now wish to propagate the
uncertainty in these input scaling relations (see Table \ref{tab:ref_relations}) 
into uncertainty in the $\Ysz$--$M$ relation.
The first step is to specify the choice of units for
each data set.  The mass units $M_0$ are set to $M_0/10^{14}\ \msun=4.8$, $5.5$, and 
$10.0$  for the V09, P11-LS, and M10 data sets.
The slope of the halo mass function $\beta$ is evaluated at this pivot point for
our fiducial cosmology, and the corresponding values are tabulated in Table \ref{tab:derived_relations}.
For the $SZ$ units, we set
$\Delta_0^2=10^{-5}\ \Mpc^2$.   We adopt the pivot point $Y_0$
quoted for the $M$--$\Yx$ relation.  Our choice of units for $M_0$ is set so as to
match the pivot point in the $\Lx$--$M$ relation, unless doing so induces strong
covariance between the amplitude and slope of the $\Ysz$--$M$ scaling relation.
This only occurs for the P11-LS(z=0.23) data set, where the mass pivot of the \citet{planck11_local}
and \citet{arnaudetal10} data are not well matched.  In this case, we chose our mass
units to decorrelate the amplitude and slope of the $\Ysz$--$M$ scaling relation, since this is the 
quantity of interest in this section.

To determine the uncertainty in the $\Ysz$--$M$ scaling relation, we randomly generate the $M$--$\Yx$
and $\Ysz$--$\Yx$ scaling
relation parameters from the values quoted in Table \ref{tab:ref_relations}.  These randomly drawn parameters
are input into equations~(\ref{eq:szm_slope}), (\ref{eq:szamp_final}) and (\ref{eq:szm_scat}) to solve for the $\Ysz$--$M$ scaling relation parameters.
Table \ref{tab:ref_relations} only quotes uncertainties for the scatter $\sigma_{m|z}$ for the M10 data set.
For the \citet{vikhlininetal09}
and \citet{arnaudetal10} $M$--$\Yx$ relation, we adopt a uniform prior on the variance $\sigma_{m|x}^2$
with $\sigma_{m|x}^2 \leq 0.07^2$  and $\sigma_{m|x}^2 \leq 0.09^2$ respectively.  
As for the M10 $\Ysz$--$\Yx$ scaling relation,  we assume a conservative flat prior on the variance
$\sigma_{sz|x}^2 \leq 0.15^2$.
We perform $10^5$ 
independent random draws to estimate the resulting likelihood distributions, also enforcing
the prior $|r|\leq 1$.
The resulting $\Ysz$--$M$ scaling relations are summarized in Table \ref{tab:derived_relations}
and Figure \ref{fig:yszm}, and the likelihood distributions are shown in Figure \ref{fig:yszm_pars}.
We emphasize that because the mass pivots are different for
different relations, one cannot use Figure \ref{fig:yszm_pars} to determine
whether the various $\Ysz$--$M$ relation are consistent with one another.  For that purpose,
one should focus instead on Figure \ref{fig:yszm}.

Finally, a warning.  The relations between the amplitudes, slopes, and scatters of the various relations
denoted above are exactly correct.  However, our Monte Carlo method for propagating the uncertainties
in the $\Ysz$--$Y_X$ and $M$--$Y_X$ scaling relations into the $\Ysz$--$M$ scaling relation
explicitly assumes that the uncertainties of these input scaling relations are uncorrelated.
However, in all
cases many of the same clusters were used as inputs in the analysis leading to the input
scaling relations were shared.  So, for instance, some of the clusters that contribute to the
$\Ysz$--$\Yx$ calibration also contribute to the $M$--$\Yx$ calibration.  In this case, the most
correct analysis would be to do a multi-variate scaling relation analysis fit from the start, something
which is not really feasible given the available data, and certainly beyond the scope
of this work.  With this caveat in mind, we proceed.


\subsection{Results}


\begin{figure}
\begin{center}
\hspace{-0.2in} \scalebox{1.2}{\plotone{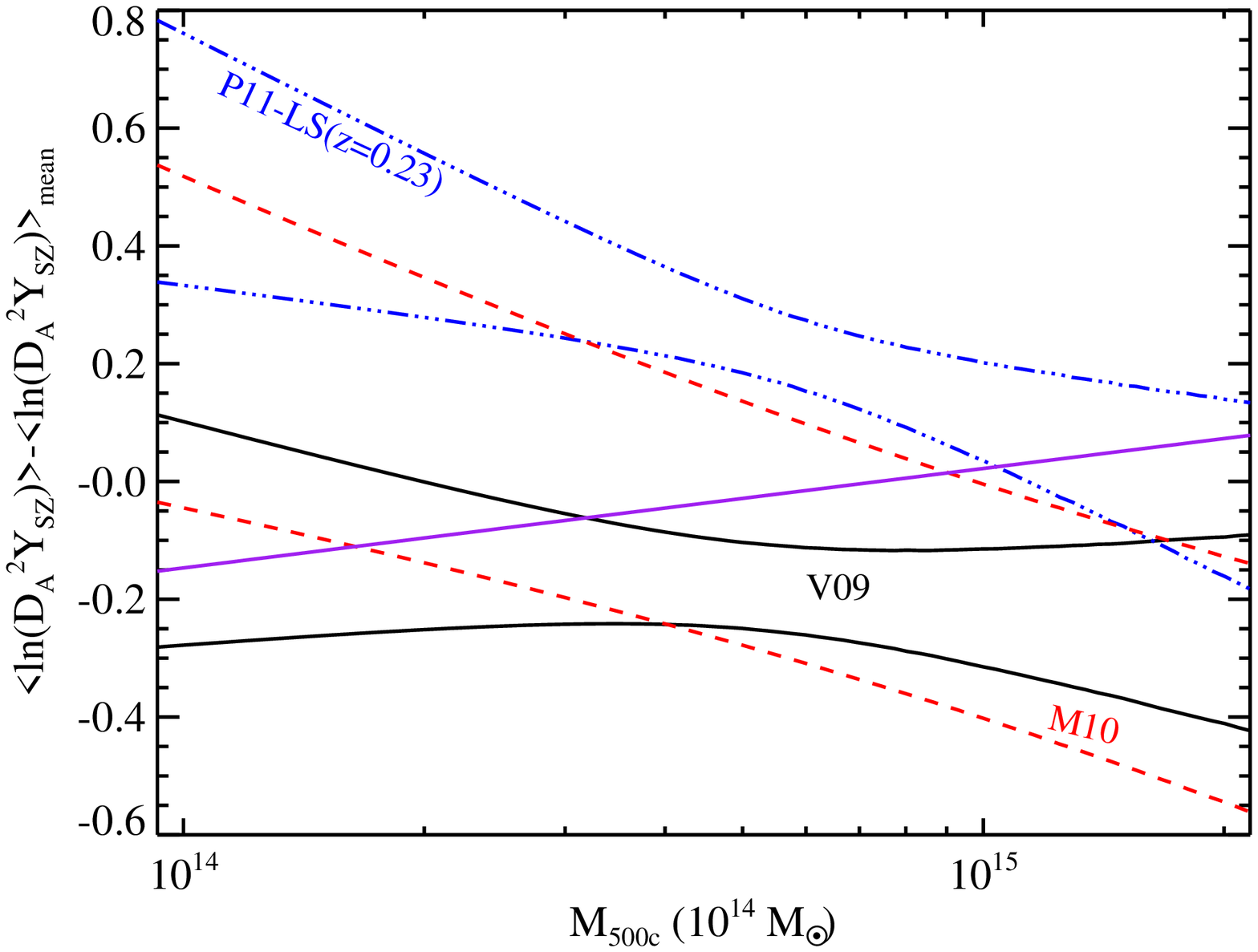}}\\
\hspace{-0.2in} \scalebox{1.2}{\plotone{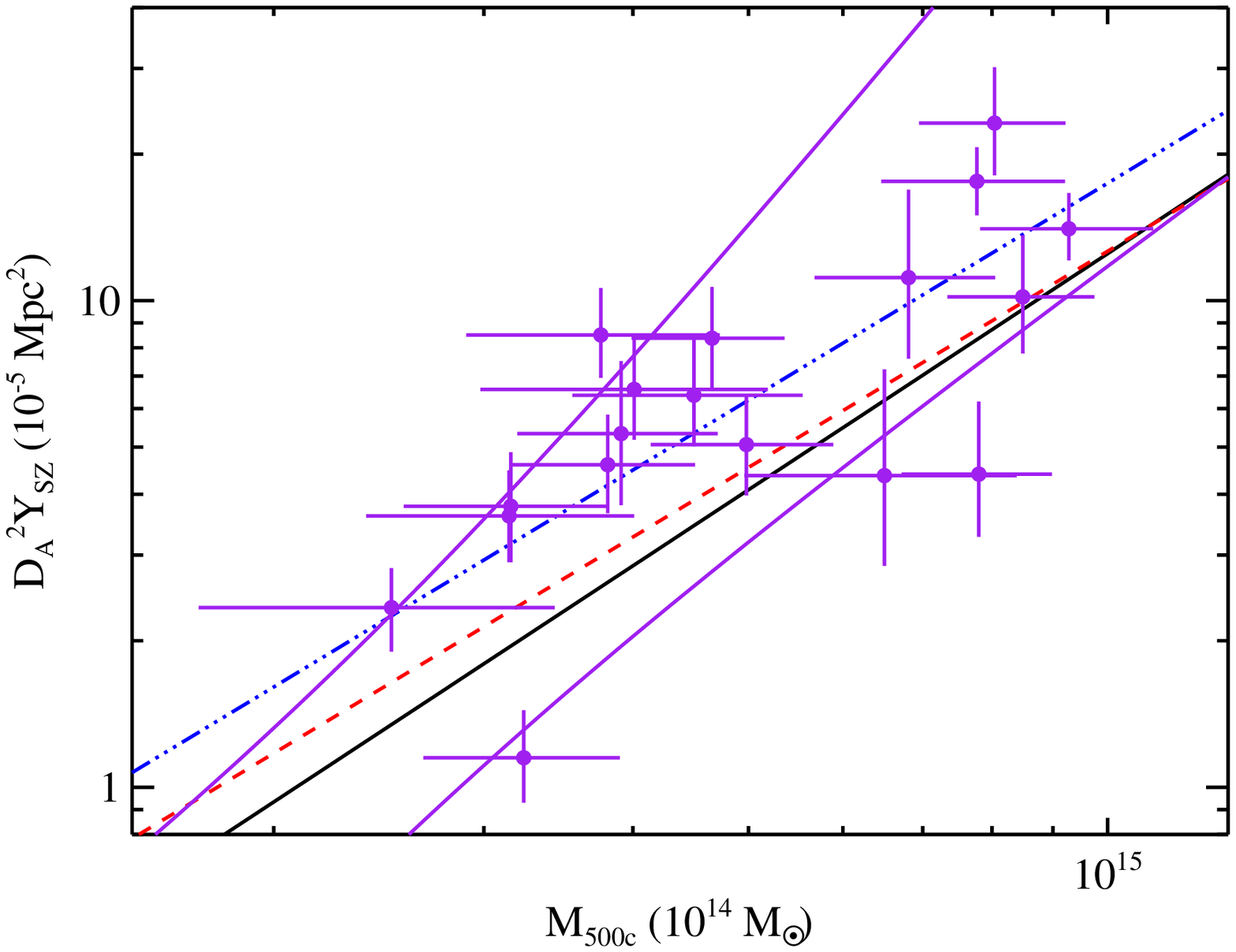}}
\caption{{\it Top panel:} $68\%$ confidence regions for the $\Ysz$--$M$ scaling relation as determined
by each of the four data sets we consider, as labeled.  The purple line is the $\Ysz$--$M$
scaling relation at $z=0.23$ as quoted in \citet{planck11_local}. {\it Bottom panel:} Comparison of our derived scaling
relations with the data from \citet{marroneetal11}.  The purple lines show the $68\%$ confidence regions
of the $M$--$\Ysz$ scaling relation from that work.  Otherwise, the color key is the same as for the top panel:
red=M10, blue=P11-LS(z=0.23), black=V09.
}
\label{fig:yszm}
\end{center}
\end{figure}


Figure \ref{fig:yszm} shows the $68\%$ confidence contours of the 
$\Ysz$--$M$ scaling relations corresponding to each of the three sets of parameters summarized
in Table \ref{tab:derived_relations}.   As usual, we subtract out a reference scaling relation,
defined by a self-similar slope of $5/3$.  The amplitude of the reference scaling relation is defined
by the mean amplitude at $M=5\times 10^{14}\ \msun$ between the V09, M10, and P11-LS(z=0.23)
data sets, and is summarized in Table \ref{tab:derived_relations}.
There is a clear amplitude offset between the V09
and P11-LS(z=0.23) data which mirrors the difference in the $\Lx$--$M$ relation.  The offset is partly
caused by the systematic differences in mass calibration between P11-LS and V09, and partly by the
redshift evolution in $\Yx$ between low and high redshifts in the P11-LS data.
The M10 relation 
is in good agreement with the V09 relation where the samples overlap ($M\sim 10^{15}\ \msun$),
but the slope of the relation is very different, reflecting the differences in slope of the $\fgas$ model.  

Comparing our self-consistently propagated scaling relation for the P11-LS(z=0.23) data set to that published
in P11-LS, we see that ours has a higher amplitude, and a different slope.  There are various reasons for this.
First, an amplitude offset is expected given the difference in the $\Ysz/\Yx$ ratio between low and high
redshift cluster samples discussed in paper I.  Likewise, a difference in the slope of the $\Ysz$--$\Yx$
relation should also result in different $\Ysz$--$M$ slopes.  The role of selection effects is a bit more unclear:
P11-LS correct for selection effects by performing Monte Carlo samples, which leads to a lowering of the amplitude
by 7\%.  On the other hand, we estimated scatter corrections to the naive plug-in method --- which we expect is roughly
equivalent to directly fitting $\Ysz$--$M$ by setting all clusters to the mass $M(\Yx)$ --- to lead to a 5\% downwards
correction, with some uncertainty associated with the value of the scatter. Because we expect that the Monte Carlo procedure
of P11-LS would detect the scatter corrections introduced in this work, it is not surprising that the two corrections are nearly
identical, and it suggests that selection function effects that are over and beyond these scatter corrections are negligible,
as would be expected for scaling relations with tight scatter.  

In an attempt to shed further light on the $\Ysz$--$M$ scaling, 
the bottom panel of Figure \ref{fig:yszm} compares our predicted scaling relations to the LoCuSS sample measurements of \citet{marroneetal11}.  That work estimated $\Mf$ using weak lensing shear measurements \citep{okabeetal10}.
The individual data points are shown, while the solid purple lines span the $68\%$ confidence region of the 
corresponding $\Ysz$--$M$ relation.  
Because
the intrinsic scatter is small, the corrections from this inversion are completely negligible relative to the statistical errors.
Consequently,
we ignored the scatter corrections when estimating the $\Ysz$--$M$ relation parameters from the published $M$--$\Ysz$
relation parameters.\footnote{We did, however, implement a prior that the slope of the $M$--$\Ysz$ relation must be larger
than $\alpha_{m|sz} \ge 0.2$.  Because the amplitude of the $\Ysz$--$M$ relation is $-a_{m|sz}/\alpha_{m|sz}$ 
(see equation \ref{eq:avgms}), allowing $\alpha_{m|sz}$ to get close to zero is both unphysical and numerically unsound.}
It is clear from the figure 
that their best fit scaling relation is in agreement with all three of the scaling relations we computed, so the LoCuSS data
is not yet of high enough quality to unambiguously prefer one data set over another.

In this context, we note that a recent paper by \citet{planck12_wl} has noted there are large systematic differences between
their X-ray masses and the weak lensing masses of \citet{okabeetal10}.  We treat this problem in more detail in paper III.  Here,
it suffices to note that these systematic differences are not sufficient to create statistical tension between our results and those
of \citet{marroneetal11} because of the large uncertainties in the latter work.


\begin{deluxetable*}{llllllll}
\tablewidth{0pt}
\tablecaption{Derived Cluster Scaling Relations at $z=0.23$}
\tablecomment{
In all cases, we assume the $\psi$--$\chi$ relation takes the form
$\ln \psi= \ln \psi_0+\alpha \ln(\chi/\chi_0)$.  Our choice of units are $10^{14}\ \msun$ for mass, 
$10^{44}\ \mbox{ergs/s}$ for $\Lx$,  and 
$10^{-5}\ \Mpc^2$ for $D_A^2\Ysz$.   All scaling relations are appropriate for 
$z=0.23$, the median redshift of the maxBCG cluster sample.  The quantity $\beta$ is the slope
of the halo mass function at the pivot scale of the $\Ysz$--$M$ relation.  All errors are the standard
deviation of the distribution, except for the correlation coefficient $r$ between $\Ysz$ and $\Yx$ at fixed mass.
The likelihood distribution for $r$ is highly
non-gaussian (see Figure \ref{fig:r}), so we quote the maximum likelihood value, and the error bars define the
$68\%$ likelihood contour. 
}
\tablehead{
Relation & $\chi_0$ & $\beta$ & $\ln \psi_0$ & $\alpha$ & $\sigma_{\ln \psi|\chi}$ & $r$ & Sample}
\startdata
$D_A^2\Ysz$--$M$ & 4.8 & $2.75$ & $1.34\pm 0.07$ & $1.61\pm 0.11$ & $0.12\pm 0.03$ & $0.75^{+0.20}_{-0.20}$ & V09 \\
$D_A^2\Ysz$--$M$ & 5.5 & $2.93$ &  $1.97\pm 0.06$ & $1.48\pm 0.12$ & $0.20\pm 0.04$ & $0.60^{+0.15}_{-0.20}$ & P11-LS(z=0.23) \\
$D_A^2\Ysz$--$M$ & 10.0 & $3.95$ &  $2.54\pm0.20$ & $1.48 \pm 0.09$ & $0.15\pm 0.03$ & $0.66^{+0.18}_{-0.09}$ & M10 \\
$D_A^2\Ysz$--$M$ & 5.0 & --- &  $1.58$ & 5/3 & --- & --- & Reference \vspace{0.05in} \\
\hline
\hline \vspace{-0.05in} \\
$D_A^2\Ysz$--$\Lx$ & 4.0 & --- &  $1.33\pm 0.13$ & $1.01\pm 0.11$ & $0.35\pm 0.07$ & --- & V09 \\
$D_A^2\Ysz$--$\Lx$ & 4.0 & --- &  $1.45\pm 0.15$ & $0.92\pm 0.10$ & $0.32\pm 0.10$ & --- & P11-LS(z=0.23) \\
$D_A^2\Ysz$--$\Lx$ & 10.0 & --- &  $2.27\pm 0.31$ & $1.10\pm 0.08$ & $0.40\pm 0.08$ & --- & M10 \vspace{0.05in}
\enddata
\label{tab:derived_relations}
\end{deluxetable*}



\section{The $\Ysz$--$\Lx$ Relation}
\label{sec:yszlx}

We now use the $\Ysz$--$M$ and $\Lx$--$M$ scaling relations to self-consistently derive the $\Ysz$--$\Lx$ scaling
relation.  There are two motivations for this work.  First, in the absence of systematic errors, our prediction for the
$\Ysz$--$\Lx$ scaling relation must be fully self-consistent with the observed $\Ysz$--$\Lx$ relation.  We have 
seen, however, that the scaling relations of galaxy clusters systematically vary from data set to data set.   If we can determine
that for a given data set our predicted $\Ysz$--$\Lx$ relation is not consistent with the observed $\Ysz$--$\Lx$
relation, then that is strong evidence that these exist systematic errors in the corresponding data set.  

To the extent that a multivariate Gaussian approximation is a valid description of the scatter about mean scalings, any data set whose relations are not self-consistent ({\sl i.e.}, deviate from the expectations of Appendix \ref{app:multi}) necessarily suffers from systematic uncertainties.
By the same token, the $\Ysz$--$\Lx$ scaling relation plays a critical self-consistency constraint for the SZ--optical scaling relations
of galaxy clusters, and will therefore provide an important test of the interpretation of the \citet{planck11_optical} results presented in paper III.

\subsection{Method}

Similar to the previous section, we define
\be
l_X = \ln \left(\Lx/L_0\right).
\ee

Having already specified the $\Ysz$--$M$ and the $\Lx$--$M$ scaling relations,
the $\Ysz$--$\Lx$ relation is constrained, up to the value of the covariance  
between $\Ysz$ and $\Lx$ at fixed $M$.  Because the scatter in
$\Lx$ is dominated by the details of the X-ray emission from the core of galaxy clusters, while the
scatter in $\Ysz$ is dominated by projection effects, the two are not likely to be correlated
at any significant level, so one might expect $r_{sz,l|m}\approx 0$.  On the other hand,
\citet{staneketal10} derived theoretical predictions for the correlation coefficient between $\Ysz$ and $L_{bol}$,
the bolometric X-ray luminosity, based on numerical simulations.  
They find $r\approx 0.8$, which reflect the fact that clusters
with more gas and/or hotter gas will have both higher X-ray luminosity and higher pressure.
In the absence of more detailed information, we adopt
a uniform prior $r_{sz,l|m} \in [0,1]$.  

Using the formulae in Appendix \ref{app:multi} \citep[see also][]{whiteetal10}, and following arguments similar to those in the previous
section, we can write the amplitude, slope, and scatter, of the $\Ysz$--$\Lx$ scaling relation in terms
of the parameters characterizing the $\Ysz$--$M$ and the $\Lx$--$M$ scaling relation.
In the notation of the previous section, we find
\bea
a_{sz|l} & = & \left[ a_{sz|m} -\frac{\alpha_{sz|m}}{\alpha_{l|m}} a_{l|m} \right] \nn \\
	& & \hspace{0.4 in} + \beta \alpha_{sz|m} \sigma_{m|l} \left[  r_{sz,l|m} \sigma_{m|sz} - \sigma_{m|l}  \right ] \label{eq:yszlx_amp} \\
\alpha_{sz|l} & = & \frac{\alpha_{sz|m}}{\alpha_{l|m}} \\
\sigma_{sz|l}^2 & = & \alpha_{sz|m}^2 \left[\sigma_{m|l}^2 + \sigma_{m|sz}^2 - 2r_{sz,l|m} \sigma_{m|l}\sigma_{m|sz} \right]
\eea
The term in square brackets is the naive relation one would obtain in the absence of scatter,
and the scatter in mass at fixed observable is related to the scatter in observable at fixed mass via equation \ref{eq:scatrel_1d}.
To get an order of magnitude estimate for the correction term, we set $\beta=3$,
$\alpha_{sz|m}=5/3$, $r_{sz,l|m}=0$, and $\sigma_{m|l}=0.25$, which yields a correction
term $\approx -0.3$, corresponding to a $30\%$ offset in $\Ysz$ at fixed $\Lx$, a very significant
correction.  If we set $r_{sz,l|m}=1$ instead, and adopt $\sigma_{m|sz}=0.1$, 
the correction is only reduced to $\approx - 0.2$, still large.

To apply this formalism, we must again specify units.  We adopt as mass units the pivot point of the
$\Ysz$--$M$ relation, which matches that of the $\Lx$--$M$ relation for the V09 and M10 data sets.
For the P11-LS data set, the pivot point of the $\Lx$--$M$ relation is lower than that of the $\Ysz$--$M$
relation.  When sampling the $\Lx$--$M$ scaling relation, we randomly draw the amplitude and slope
at the pivot point of the $\Lx$--$M$ relation, and then shift the amplitude to the new pivot point.  This procedure
explicitly includes the induced covariance between amplitude and slope at the new pivot point.  
The scaling relations are sampled $10^{4}$ times, and the corresponding $\Ysz$--$\Lx$ parameters
constrained from the Monte Carlo distribution.  The units for $\Lx$ are chosen so as to decorrelate
the amplitude and slope of the $\Ysz$--$\Lx$ relation, and are set to $L_0=3\times 10^{44}\ \ergs/\mbox{s}$
for the V09 and P11-LS data sets, and $\Lx=10^{45}\ \ergs/\mbox{s}$ for the M10 data set.
The units for $D_A^2\Ysz$ are set to $\Delta_0^2 = 10^{-5}\ \Mpc^2$ in all cases.
As in the case of the $\Ysz$--$M$ relation, we note that in principle the uncertainties in the input scaling relation
parameters can be correlated, for instance due to shared clusters in the calibration of the $\Lx$--$M$ and $\Ysz$--$M$
scaling relations, so the same caveat mentioned in the previous section holds for this discussion as well.


\begin{figure}
\begin{center}
\hspace{-0.2in} \scalebox{1.2}{\plotone{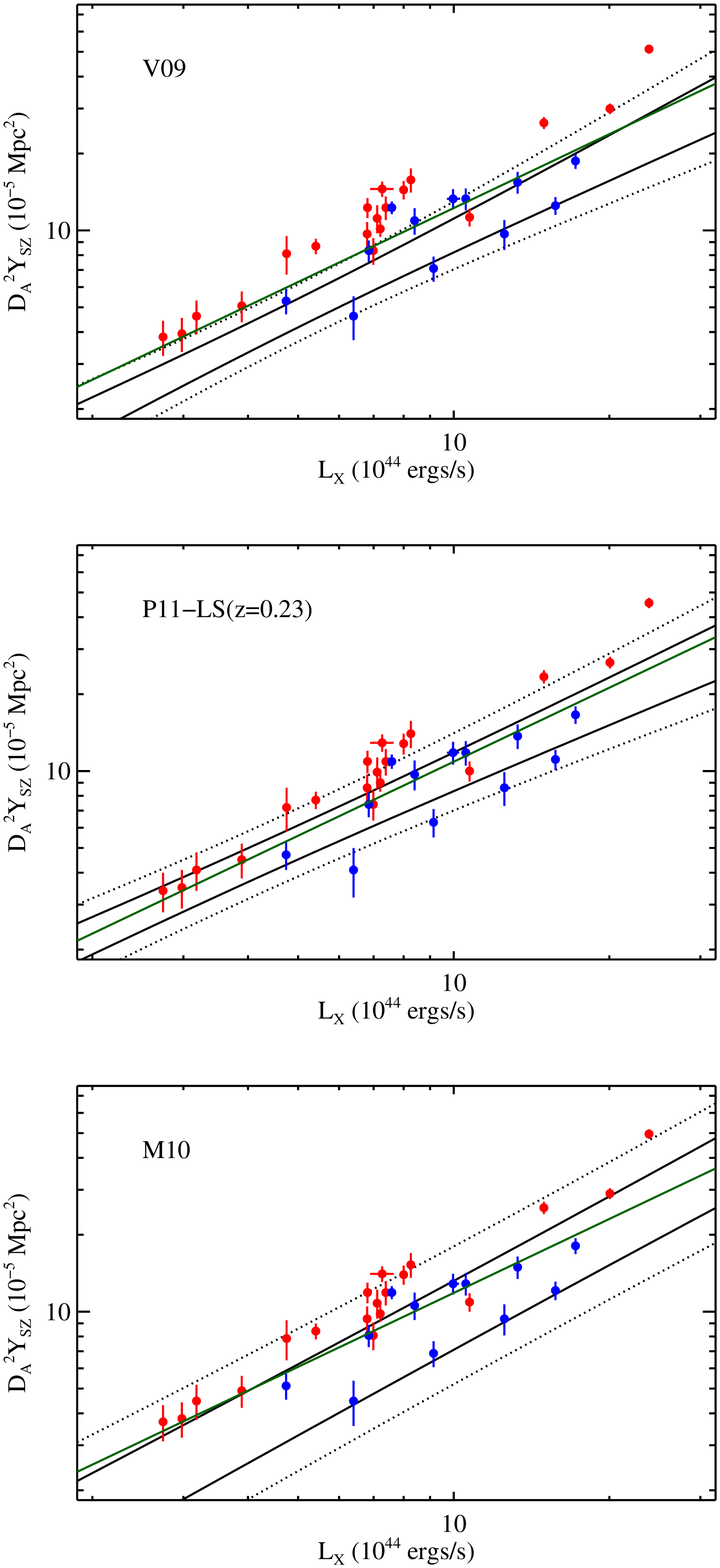}}
\caption{Comparison between the \citet{planck11_local} data for clusters in the redshift range $z\in[0.12,0.3]$ 
and  the $\Ysz$--$\Lx$  relations derived 
self-consistently from the $\Ysz$--$M$ and $\Lx$--$M$ relations.
The green line is our best fit to the \citet{planck11_local} data, while the black solid (dotted) lines mark the
$1\sigma$ $(2\sigma)$ confidence regions of the mean scaling relation.  Blue points are cool-core systems,
while red points are not. 
}
\label{fig:yszlx_wdata}
\end{center}
\end{figure}



\subsection{Results}

Figure \ref{fig:yszlx_pars} shows the likelihood for each of the data sets we consider.
There are significant differences in amplitude between the various works, but the uncertainties are also large.   
We compare each of our predictions to the \citet{planck11_local} data. Figure \ref{fig:yszlx_wdata} shows all galaxy clusters
in \citet{planck11_local} in the redshift slice $[0.13,0.3]$.  We adopt this cut to ensure that we only look at clusters near the
redshift of interest ($z=0.23$), and also because from paper I we know that the low and high redshift populations of the \citet{planck11_local}
clusters appear to have different properties.   
Note that because $\Ysz$ depends on $\Rf$, when comparing our predictions to the P11-LS $\Ysz$ data we need to account for
the systematic difference in mass between the various groups as per the Appendix B in paper I.  
We do so by shifting the data points rather than our predictions, though we also note that because $\Ysz$
depends only mildly on $\Rf$, the shifts due to aperture corrections are much smaller than the differences between the predicted
scaling relations (i.e. aperture corrections in $\Ysz$ are a second order effect).

In addition to showing the data points themselves, Figure \ref{fig:yszlx_wdata} shows as a dark-green line
our best-fit relation to the P11-LS(z=0.23) data.  The amplitude of the fit is corrected for selection 
effects based on the results of P11-LS,
which amount to a 9\% reduction of the best fit amplitude.
The $1\sigma$ and $2\sigma$ regions for each
of the three
predicted scaling relations are shown in Figure \ref{fig:yszlx_wdata} as black solid and dotted lines respectively.
The best fit relation to the data is clearly that derived from the P11-LS(z=0.23) scaling relations, but  
the best fit relation is within the $2\sigma$ regions
of the predicted scaling relations for all three data sets.  Consequently, we are unable to identify
which data sets suffer from systematic errors using the $\Ysz$--$\Lx$ measurements from \citet{planck11_local}.

\begin{figure}
\begin{center}
\hspace{-0.0in} \scalebox{1.2}{\plotone{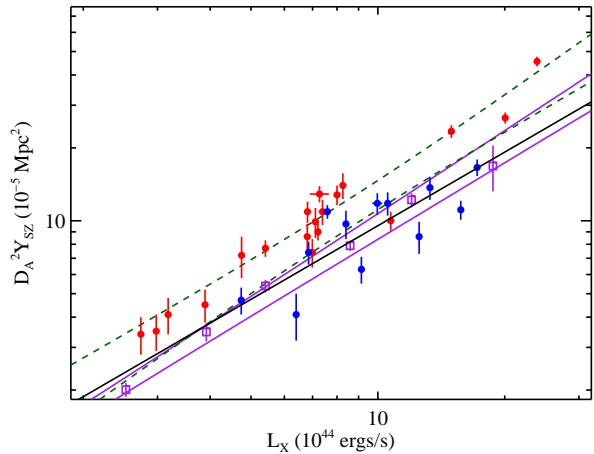}}
\caption{Comparison between the P11-LS (dark-green dashed lines) and P11-X (purple solid
lines) fits for the $\Ysz$--$\Lx$ relation at $z=0.23$.  Both bands represent $2\sigma$ confidence intervals.  
Blue points are cool-core systems,
while red points are not.  At $2\sigma$,
the two regions barely overlap.  The difference in amplitude at  the pivot point of the P11-LS data ($\Lx=7\times 10^{44}\ \ergs/\mbox{s}$)
is $\Delta \ln \Ysz = 0.29$,
which is significant at the $3.2\sigma$ level.   The black solid line is the predicted $\Ysz$--$L_X$ relation from the V09 scaling
relations, and is indicative of the excellent agreement between the predicted $\Ysz$--$L_X$ relations and the P11-X results.
The agreement of P11-X with M10 is also excellent.
}
\label{fig:yszlx_fits}
\end{center}
\end{figure}

We have also attempted to compare our self-consistently propagated scaling relations to the data 
from \citet[][]{planck11_xray}, henceforth referred to as P11-X.
P11-X assembled a heterogenous X-ray cluster catalog, the MCXC catalog \citep{piffarettietal11},
and used it to measure the $\Ysz$--$\Lx$ relation.  Unfortunately, this comparison is not trivial.
As noted in that work, the $\Ysz$--$\Lx$ scaling relation from the MCXC sample and the P11-LS scaling
relations are not consistent, with the MCXC relation shifted towards more luminous systems (lower amplitude).
This is illustrated in Figure \ref{fig:yszlx_fits}, which show the $2\sigma$ regions of the P11-X results
as the solid purple lines.\footnote{To compute these regions, we used the fits from Table 4 in P11-X 
where both the amplitude and slope of the relation are allowed to vary.  We also 
 corrected to account for the difference between $\avg{\Ysz|\Lx}$ --- the quantity measured by P11-X ---
and $\avg{\ln \Ysz|\Lx}$.}
Also shown as purple squares with errors are the binned data from
the same work.
The dashed dark-green lines are the $2\sigma$ confidence regions as reported
in Table 2 of \citet[][i.e. the P11-LS result]{planck11_local}.  
The $z\in[0.12,0.3]$ clusters in P11-LS are shown as circles with error bars,
and our own fit to the data is similar to that of P11-LS.
We see that the $2\sigma$ boundaries of the two relations --- that of P11-LS and that of \citet{planck11_xray} --- 
barely overlap.  The
amplitude difference at the pivot scale of the P11-LS data ($\Lx=7\times 10^{44}\ \msun$)
is $\Delta \ln \Ysz = 0.29$, a $3.2\sigma$ offset.  

P11-X attributes the offset in their result relative to that of P11-LS to selection effects in the MCXC sample.  
However, we believe this interpretation is not correct
for several reasons.    First, the P11-X sample relied on archival data, which heavily favors X-ray
selected systems.  Thus, it is unclear whether the MCXC catalog is really any more ``X-ray selected'' than the P11-LS sample.
More importantly, we emphasize 
that we are considering the $\Ysz$--$\Lx$ relation: that is, $\Ysz$ is consider the dependent variable.  
By definition, if our sample contains all galaxy clusters of a given
X-ray luminosity, we can explicitly compute $\avg{\Ysz|L_X}$ without any corrections, i.e. there are no selection effects of which
to speak of.
Finally, P11-X argue that the reason for the offset is a larger presence of cool-core systems in
their data relative to P11-LS.   We have refit the $z\in[0.13,0.3]$ data using only cool-core systems.  The resulting best
fit relation is still slightly to the left of the \citet{planck11_xray} relation, with an amplitude offset  $\Delta \ln \Ysz=0.08\pm 0.14$.
This is consistent with zero, but taken at face value, would argue that {\it all} galaxy clusters in \citet{planck11_xray} must
be cool-core systems in order to account for the observed offset.  


Regardless of these difficulties, we can still compare the P11-X scaling relation to our
predicted V09, M10, and P11-LS(z=0.23) $\Ysz$--$L_X$ relations.  As a simple illustrative example,
Figure \ref{fig:yszlx_fits} shows as a black line the V09 prediction for the $\Ysz$--$L_X$ scaling relation.
In all cases, we find that the P11-X
result is in better agreement with our predictions than the raw P11-LS data.

In short, it is not clear what drives the difference between the $\Ysz$--$\Lx$ relations of P11-X
and P11-LS, nor which $\Ysz$--$L_X$ scaling relations are ultimately correct.  
The best we can say is that all three predicted $\Ysz$--$\Lx$ scaling relations --- V09, M10, and P11-LS(z=0.23) --- 
are viable, with better agreement found between our predictions and the P11-X results.
We will return to the problem of the $\Ysz$--$L_X$ relation in paper III.


\section{Cosmological Consistency}
\label{sec:cosmology}

Having performed a detailed comparison of the cluster scaling relations from three different
data sets, we test which of these relations are also consistent with
the currently favored WMAP7 $\Lambda$CDM cosmological model.  This test is not
a priori: because of the cosmological analysis in \citet{vikhlininetal09b} and
\citet{mantzetal10a}, we already know that both the V09 and M10 scaling relations are consistent
with WMAP7.  Here, we confirm this agreement, and test the additional $L_X$--$M$ scaling relation
of \citet{prattetal09}, which uses the same mass calibration as P11-LS.

We proceed as follows.  We randomly sample the CMB+BAO+$H_0$ Monte Carlo Markov Chains (MCMC)
from \citet{komatsuetal11} for a ``vanilla'' flat $\Lambda$CDM cosmology.  The parameters that we vary are
$\sigma_8$, $n_s$, $\Omega_m$, $h$, and $\Omega_b$.
For each random sampling, we compute the \citet{tinkeretal08} mass function.
We also randomly sample the amplitude, slope, and scatter of the $L_X$--$M$ relation 
for each of the V09, M10, and P11-LS scaling relations, and convolve the resulting
mass function and $P(L_X|M)$ relations to arrive at the predicted X-ray luminosity function,
\be
\frac{dn}{d \ln L} = \int d(\ln M)\ \frac{dn}{d\ln M} P(\ln L|\ln M).
\ee
The uncertainty in the predicted X-ray luminosity function is estimated from the variance over our Monte Carlo
realizations.

We compare this prediction to the X-ray luminosity function from the REFLEX cluster survey \citep{bohringeretal02}.
Note that both the reported $P(\Lx|M)$ scaling relations and the REFLEX luminosity function measurement assume $\Omega_m=0.3$
rather than the WMAP7 central value $\Omega_m=0.27$.  We do not expect this difference to be an important 
systematic.  For instance, the difference in the luminosity distance out to $z=0.23$
between the two models is only $\approx 0.5\%$, so we do not expect more than a few percent changes over the range of cosmologies
considered here.  These differences are much smaller than the relative differences between the $\Lx$--$M$ relations considered
here.

\begin{figure}
\begin{center}
\hspace{-0.0in} \scalebox{1.2}{\plotone{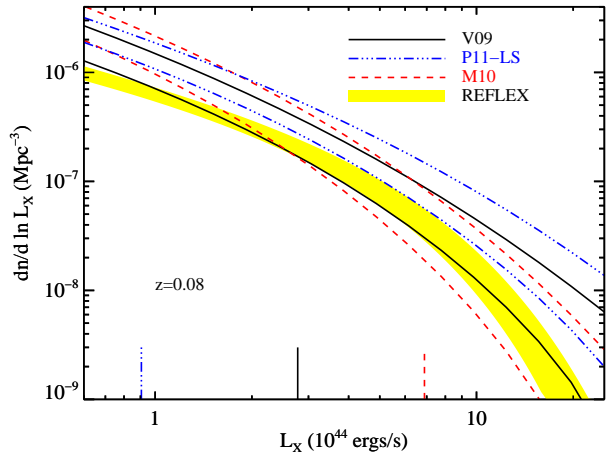}}
\caption{Comparison of the X-ray luminosity predicted from the $\Lx$--$M$ relation
derived from each of the data sets to the REFLEX luminosity function, as labelled.  The bands shown
correspond to $68\%$ confidence, and predictions
assume a WMAP7 cosmology \citep{komatsuetal11} sampled over $\sigma_8$, $\Omega_m$, $h$, $n_s$,
and $\Omega_b$. 
The vertical lines along the $x$-axis mark
the pivot point of each $\Lx$--$M$  relation.   Both V09 and M10 are consistent with cosmological expectations,
while the P11-LS relation is in mild tension ($2.1\sigma$) with WMAP7.  Adding BOSS priors (see text)
increases the tension with P11-LS to $3.3\sigma$, while both V09 and M10 remain consistent with the REFLEX
luminosity function. 
Both the mass function and the scaling relations are
evaluated at $z=0.08$, the median redshift of the REFLEX catalog.  
}
\label{fig:counts}
\end{center}
\end{figure}

Figure \ref{fig:counts} shows the comparison between our predicted X-ray luminosity functions, and the REFLEX luminosity function.
For our predictions, we evaluate both the mass function and  $\Lx$--$M$ scaling relations at $z=0.08$, the median redshift
of the REFLEX clusters.  The dependence of our results on the assumed redshift is mild.
The bands for our theoretical predictions show the $68\%$ confidence intervals,
estimated from $10^3$ random samplings of the cosmological and scaling relation parameters.
The width of the REFLEX band
is computed by random sampling of the Schechter luminosity function parameters 
$L_*$ and $\alpha$ using the quoted uncertainties in \citet{bohringeretal02}.
In addition, it is important to note that while in our work $\Lx$ was defined using an aperture $\Rf$, the REFLEX catalog defines
$\Lx$ using growth curve analysis, so the two luminosities are not directly comparable.  Because $\Lx$ varies very slowly
with aperture, we expect that a $\pm 10\%$ systematic uncertainty in the luminosity is a reasonable estimate the impact that this
difference can have.  We convert this uncertainty into an uncertainty in the luminosity function using the logarithmic derivative of
the average luminosity function, and add this in quadrature to the statistical error to arrive a the total uncertainty in the empirical
luminosity function.


The small vertical lines along the $x$-axis denote the $\Lx$ corresponding to the mass
pivot point of each of the three $\Lx$--$M$ relations.  For properly calibrated scaling relations, the predicted
abundance function should be statistically consistent the observed abundance at the pivot point.
As expected, the V09 and M10 models are in good agreement with WMAP7, with the observed offsets significant 
at $0.6\sigma$ and $0.3\sigma$ respectively.  The P11-LS model is in mild tension ($2.1\sigma$) with WMAP7.

We have repeated this experiment using the additional cosmological priors from the Baryon Oscillation Spectroscopic
Survey (BOSS) experiment \citep{sanchezetal12}.  Because the full likelihood is not yet published, we sample only the
cosmological parameter $\sigma_8\Omega_m^{1/2}$, corresponding to the standard cluster normalization condition.
The cosmological prior from BOSS for this parameter is $\sigma_8\Omega_m^{1/2}=0.441\pm 0.013$ (A. Sanchez, private
communication).  All other cosmological
parameters are held fixed to their fiducial value.  We have further verified that when we perform this simplified analysis on the
WMAP7 results, our results are very nearly identical to the results when we allow all cosmological parameters to vary.  
The addition of BOSS data sharpens the previous discussion: V09 and M10 remain consistent with WMAP7+BOSS,
with the abundance offset being significant at $1.5\sigma$ and $0.7\sigma$ respectively.  The offset relative to the
P11-LS prediction is $3.3\sigma$, in modest tension with the WMAP7+BOSS results.

Interestingly, an examination of Figure \ref{fig:lxm} ($\Lx$--$M$), Figure \ref{fig:myx} ($M$--$\Yx$),
Figure \ref{fig:yszyx} ($\Ysz$--$\Yx$), Figure \ref{fig:yszm} ($\Ysz$--$M$), and Figure \ref{fig:yszlx_wdata} ($\Ysz$--$\Lx)$,
reveal
that at the pivot point of the M10
data set, the amplitude of the V09 scaling relations are in good agreement with the amplitude of the M10 scaling relations.
That is, the V09 and M10 relations are in good agreement for the most massive clusters,
but diverge at low masses because of the differences in the $\fgas$--$M$ relation between the two works.  
From Figure \ref{fig:counts}, we see
that the $\fgas$ model of V09 is a better match to the full set of scales probed by REFLEX.  
This stands is in contrast to the results of \citet{allenetal08}.  However, the latter work focused on the most
massive galaxy clusters in the Universe, so the two results may be reconciled if there is a flattening of the slope in the 
$\fgas$--$M$ relation at the highest masses.


\section{Summary and Conclusions}
\label{sec:summary}

We have performed a detailed comparison of the X-ray scaling relations from three different data sets:
P11-LS \citep{prattetal09,arnaudetal10,planck11_local}, 
V09 \citep{vikhlininetal09,rozoetal12a}, and M10 \citep{mantzetal10b}.
We compare the $\Lx$--$M$, $M$--$\Yx$, and $\Ysz$--$\Yx$ scaling relations and find varying degrees of tension between the works.   Differences are partly traced to the systematic offsets in cluster observables characterized in paper I, but the constant $\fgas$ model of M10 drives slope differences relative to V09 and P11-LS.  After correcting for these two effects, all cluster scaling relations are in good agreement
with each other.  Indeed, at the pivot point of the M10 scaling relations, all of the V09 and M10 
scaling relations are in good agreement with each other: the difference between these two works is pronounced only if one extrapolates the M10 scaling relations to low mass systems.  

Having identified the sources of tension, we use these scaling relations to self-consistently recover the $\Ysz$--$M$ and $\Ysz$--$\Lx$ 
scaling relations within the context of a multivariate Gaussian model for cluster properties \citep[see Appendix \ref{app:multi}, also][]{whiteetal10}.   
In the limit of very small scatter, the transfer relations are equivalent to ``plugging-in'' one relation into 
another.  In the presence of finite scatter, there are additional terms involving property covariance and the local slope of the 
mass function that must be taken into account.  For instance, in the case of the $\Ysz$--$\Lx$ scaling relation, these corrections 
are of order $25\%$.  

The differences in mass calibration between the various data sets, presented in paper I,  lead to different predictions for the $\Ysz$--$M$ 
scaling relation.  Independent results from the LoCuSS collaboration \citep{marroneetal11} are not yet able to distinguish between the 
various predictions.

For $\Ysz$--$\Lx$ scaling, we find that the differences between the various data sets are 
moderate.   Indeed, we compared each of the predicted $\Ysz$--$\Lx$
scaling relations to the P11-LS data, and found reasonable agreement in all cases.  
We also compared our predicted
$\Ysz$--$\Lx$ scaling relations with those of \citet[][referred to as P11-X]{planck11_xray}, noting that there are large differences
between the P11-LS and the P11-X results ($4.1\sigma$ significance), finding much better agreement with
the P11-X results for the V09 and M10 data sets.
The selection effects required to reconcile the P11-LS and P11-X would have to be large, with the MCXC catalog comprised
exclusively of cool-core clusters, or $SZ$ selection leading to an observed $30\%$ amplitude shift; either solution seems 
too large to be plausible, as 
correction terms from selection effects in SZ typically scale as the variance $\sigma_{sz|m}^2 \approx 0.01$, where as the observed
offset is nearly 30\% in $\Ysz$.

Finally, we consider whether the scaling relations from each of the three data sets are consistent with cosmological expectations.   
We convolve the three $\Lx$--$M$ relations with the mass function predicted by a WMAP7 cosmology --- sampling over $\sigma_8$,
$\Omega_m$, $n_s$, $h$, and $\Omega_b$ ---
and compare these predictions against the observed REFLEX luminosity function 
\citep{bohringeretal02}.  As expected, the M10 and V09 scaling relations are consistent with the REFLEX luminosity function at the 
pivot point of the two samples.   The prediction from the  P11-LS $\Lx$--$M$ scaling relation is offset from the REFLEX data by $2.1\sigma$.  
Adding BOSS priors increases the tension with P11-LS to $3.3\sigma$, while both the V09 and M10 predictions remain consistent with 
the data.
In addition, comparing the M10 and V09 predictions, it is clear that the abundance of low mass systems is
correctly predicted by V09 but not by M10, which argues that an $\fgas$ model $\fgas\propto M^\gamma$
with $\gamma\approx 0.10-0.15$ is a better match to the data over the full range of scales probed by REFLEX galaxy clusters.  
Because $\fgas$ is observed to be constant at high masses \citep{allenetal08}, this may be
signaling a steepening of the $\fgas$--$M$ relation at low masses.

The fact that the three different X-ray data sets considered here and in paper I exhibit systematic differences in mass calibration
and cluster scaling relations has important consequences for the predictions of the $\Ysz$--$N_{200}$ scaling relation
for optical galaxy clusters in \citet{planck11_optical}.  In particular, the fact that the different data sets 
give rise to different $\Ysz$--$M$ scaling relations also implies that the predicted $\Ysz$--$N_{200}$ relations
from each of these data sets will be different.  In paper III, we explore whether any of these
three data sets can resolve this issue, and whether doing so
results in a self-consistent picture of multi-variate cluster scaling relations.  More specifically, we will will demand
not just that the predicted $\Ysz$--$N_{200}$  relation match observations, but also that any two scaling relations can 
be combined to successfully predict the third, while simultaneously satisfying cosmological expectations for the
counts of galaxy clusters selected by any property.

\acknowledgements The authors would like to thank Adam Mantz, Alexey Vikhlinin, Gabriel Pratt, Monique Arnaud, 
and Steven Allen for useful criticisms 
on earlier drafts of this work.  The authors would also like to thank the organizers of the Monsters Inc. workshop at KITP, 
supported in part by the National Science Foundation under Grant No. PHY05-51164,
where this collaboration was started. 
ER gratefully acknowledges the hospitality of the AstroParticle and Cosmology 
laboratory (APC) at the Universit\'e Paris Diderot, where part of this work took place. 
ER is funded by NASA through the Einstein Fellowship Program, grant PF9-00068.  
AEE acknowledges support from NSF AST-0708150 and NASA NNX07AN58G.  
JGB gratefully acknowledges support from the Institut Universitaire de France.
A portion of the research described in this paper was carried
out at the Jet Propulsion Laboratory, California Institute of Technology, under a
contract with the National Aeronautics and Space Administration.
This work was supported in part by the U.S. Department of Energy contract to SLAC no. DE-AC02-76SF00515.

\newcommand\AAA[3]{{A\& A} {\bf #1}, #2 (#3)}
\newcommand\PhysRep[3]{{Physics Reports} {\bf #1}, #2 (#3)}
\newcommand\ApJ[3]{ {ApJ} {\bf #1}, #2 (#3) }
\newcommand\PhysRevD[3]{ {Phys. Rev. D} {\bf #1}, #2 (#3) }
\newcommand\PhysRevLet[3]{ {Physics Review Letters} {\bf #1}, #2 (#3) }
\newcommand\MNRAS[3]{{MNRAS} {\bf #1}, #2 (#3)}
\newcommand\PhysLet[3]{{Physics Letters} {\bf B#1}, #2 (#3)}
\newcommand\AJ[3]{ {AJ} {\bf #1}, #2 (#3) }
\newcommand\aph{astro-ph/}
\newcommand\AREVAA[3]{{Ann. Rev. A.\& A.} {\bf #1}, #2 (#3)}

\bibliographystyle{apj}
\bibliography{mybib}

\appendix


\section{A Local Model for Multi-Variate Scaling Relations}
\label{app:multi}

We consider the problem of a locally power-law halo mass function $dn/d\ln M \propto M^{-\beta}$,
and a vector of observable signals $\bmm{S}$, e.g. $\bmm{S}=\{\Lx,\Yx,\Mgas,\Ysz, etc.\}$.
The results summarized here are an extension of those presented in \citet{staneketal10} and \citet{allenetal11},
and are basically equivalent to those in Appendix C of \citet{whiteetal10}.
We  define $\mu = \ln(M/M_0)$ and $s=\ln(S_i/S_{i,ref})$ where $M_0$ and $S_{i,ref}$ represent a choice of 
units.  The scaling relations between $\bmm{S}$ and $M$ are governed by
the probability distribution $P(\bs|m)$ which we assume to be Gaussian.
The means of the distributions are parameterized as
\be
\avg{\bs|\mu} = \bmm{a} + \bm{\alpha} \, \mu.
\ee
The scatter is characterized by a covariance matrix, $C$, which has the property variance, $\sigma_i^2$, along the diagonals and off-diagonal terms, $C_{ij} = \avg{ (s_i - \avg{s_i}) (s_j - \avg{s_j}) }$.  The correlation coefficient, given by $r_{ij} = C_{ij}/\sigma_i\sigma_j$, lies between $-1$ and $1$.

Using Bayes Theorem, we can relate the $\bS$--$M$ scaling relation to the $M$--$\bS$ scaling relation
\be
P(\mu|\bs) = \frac{P(\bs|\mu)P(\mu)}{P(\bs)} = \frac{P(\bs|\mu)P(\mu)}{\int d\mu\ P(\mu)P(\bs|\mu)}.
\ee
For a locally power-law model, $P(\mu)\propto \exp(-\beta \mu)$, the resulting probability density is Gaussian 
with mean and variance
\bea
\avg{\mu|\bs} & = &  \balpha \bC^{-1} (\bs-\bm{a}) \sigma_{\mu|\bs}^2 - \beta \sigma_{\mu|\bs}^2 \\
\sigma_{\mu|\bs}^2 & = & \left(\balpha \bC^{-1} \balpha\right)^{-1}
\eea
For a single property, $s$, the above expressions reduce to
\bea
\avg{m|s} & = &  \frac{s-a}{\alpha} - \beta \sigma_{\mu}^2 ,  \label{eq:avgms}  \\
\sigma_{\mu} & = & \frac{1}{\alpha}\sigma_s.  \label{eq:scatrel_1d} 
\eea
The correction term to the mean mass, $-\beta\sigma_{\mu}^2$, is the standard Malmquist bias correction. Here, $\sigma_{\mu}$ is the scatter in halo mass of a sample selected by property $s$. 

The space density of clusters with multiple properties, $\bmm{s}$, is
\be
n(\bs) = \int d\mu\ P(\bs|\mu)n(\mu) = \frac{A\sigma_{m|\bs}}{(2\pi)^{(N-1)/2}|\bC|^{1/2}}\exp\left[ - \frac{1}{2}\left( (\bs-\bm{a})\bC^{-1}(\bs-\bm{a}) 
	- \frac{\avg{m|\bs}}{\sigma_{\mu|\bs}^2} \right) \right].
\ee
In the case of a single signal $s$, this reduces to
\be
n(s) = \frac{A}{\alpha} \exp\left[ - \beta \left( \frac{s-a}{\alpha} - \frac{1}{2}\beta \sigma_{\mu}^2 \right) \right].
\ee
Note the slope of the abundance function is simply $\beta/\alpha$, as we would expect.  There is a correction term to the amplitude, $\beta^2\sigma^2$, which reflects the number boost due to low mass halos scattered upward.

Of particular interest to us is the case in which there are two signals $s_1$ and $s_2$.  Let $\sigmuone$ and $\sigmutwo$ be the mass scatter in each, respectively.   Then the mean selected mass, $\avg{\mu|\bs}$, and the mass variance for joint signals, $s_1$ and $s_2$, are 
\bea
\avg{\mu|s_1,s_2} & = & \frac{ \sigmuone^{-2} \alpha_1^{-1} (s_1-a_1) + \sigmutwo^{-2} \alpha_2^{-1} (s_2-a_2) 	- r\sigmuone^{-1}\sigmutwo^{-1}( \alpha_1^{-1} (s_1-a_1) + \alpha_2^{-1} (s_2 - a_2)) }
	    { \sigmuone^{-2} + \sigmutwo^{-2}	- 2r\sigmuone^{-1}\sigmutwo^{-1} } \\
\sigma_{\mu|\bs}^2 & = & \frac{1-r^2}{\sigmuone^{-2} + \sigmutwo^{-2} - 2r\sigmuone^{-1}\sigmutwo^{-1} }.
\eea

Finally, a further application of Bayes theorem allows us estimate $P(s_2|s_1)$,
\be
P(s_2|s_1) = P(s_1,s_2)\frac{1}{P(s_1)} = \frac{1}{P(s_1)}\int dm\ P(s_1,s_2|\mu) P(\mu)
\ee
from which we find that $P(s_2|s_1)$ is a Gaussian distribution with mean and variance
\bea
\avg{s_2|s_1} & = & a_{2|m} + \alpha_2 \left( \avg{\mu|s_1}  + r \beta \sigmuone \sigmutwo \right)  \label{eq:avgs2s1} \\
\sigma_{2|1}^2 & = & \alpha_2^2\left[ \sigmuone^2 + \sigmutwo^2 - 2r \sigmuone \sigmutwo \right] \label{eq:scatrel}
\eea
The correlation coefficient between $s_2$ and $m$ at fixed $s_1$ is given by
\be
r_{2,\mu|1} = \frac{ \sigmuone / \sigmutwo - r }{ \left[ 1 - r^2 + (\sigmuone / \sigmutwo-r)^2 \right]^{1/2} }.
\ee


\bigskip
\bigskip


\section{Likelihood Plots}

\begin{figure*}[h!]
\begin{center}
\hspace{-0.0in} \scalebox{1.0}{\plotone{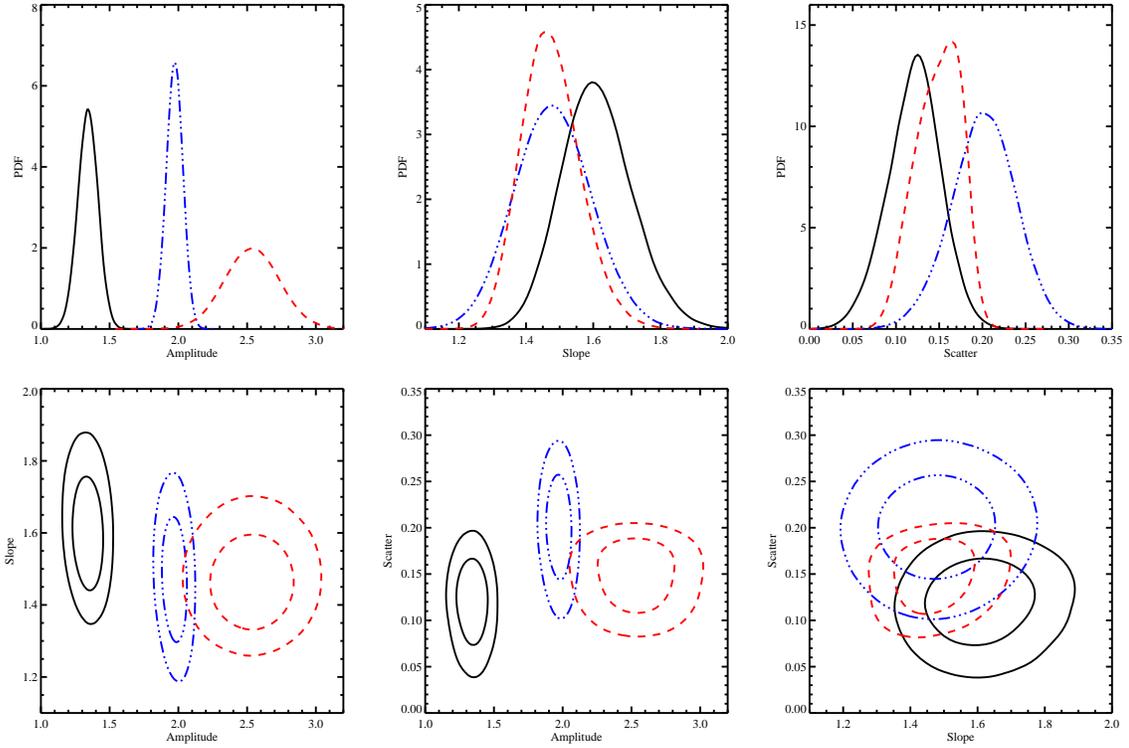}}
\caption{Likelihood distribution and likelihood contours for the parameters of the $\Ysz$--$M$
scaling relation.  Black solid=V09, blue dash--dot=P11-LS(z=0.23), and
red dashed=M10.   All contours are $68\%$ and $95\%$ confidence.
Note the different scaling relation assume different pivot points, so one cannot directly compare the amplitudes
quoted between all data sets.
}
\label{fig:yszm_pars}
\end{center}
\end{figure*}

\begin{figure*}
\begin{center}
\hspace{-0.0in} \scalebox{1.0}{\plotone{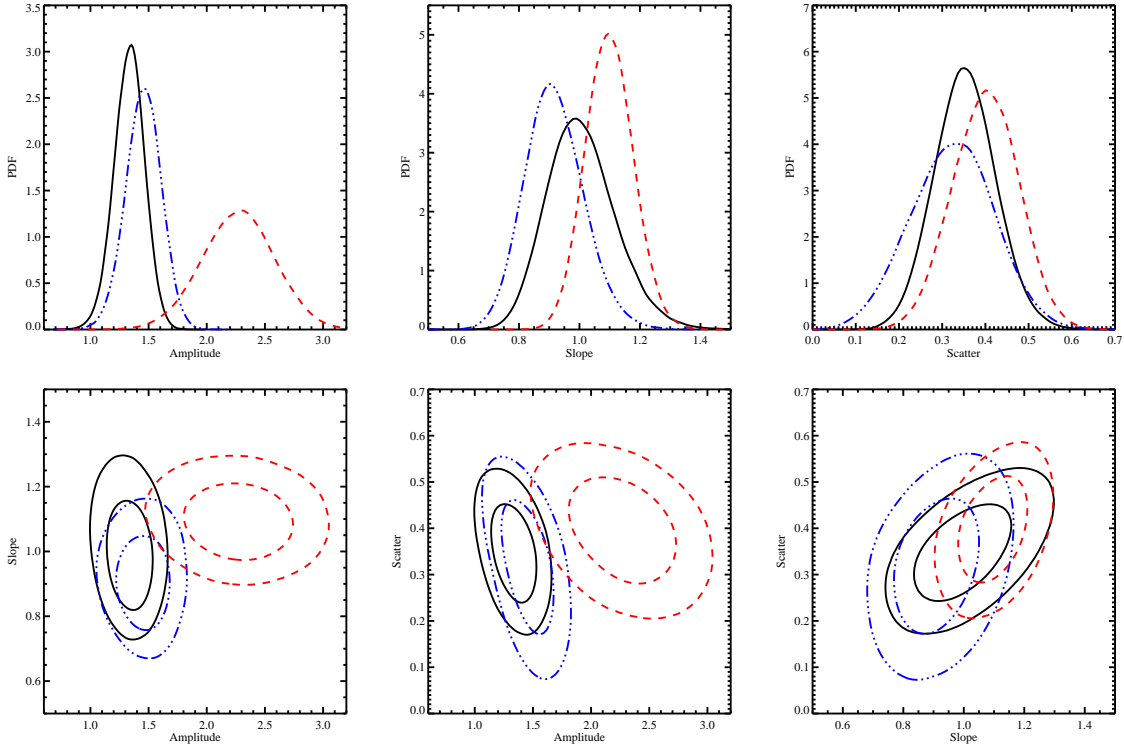}}
\caption{As Figure \ref{fig:yszm_pars}, but for the $\Ysz$--$\Lx$ parameters.
}
\label{fig:yszlx_pars}
\end{center}
\end{figure*}

Figures \ref{fig:yszm_pars} and \ref{fig:yszlx_pars} show the posterior distribution for the $\Ysz$--$M$ and $\Ysz$--$\Lx$ scaling relation parameters
summarized in Table \ref{tab:derived_relations}.


\end{document}